\documentclass[prb,twocolumn,showpacs,amsmath,amssymb,superscriptaddress,longbibliography]{revtex4}

\usepackage{graphicx}
\usepackage{dcolumn}
\usepackage{bm}
\usepackage{gensymb}
\usepackage{multirow}
\usepackage[none]{hyphenat}
\usepackage{epstopdf}
\usepackage{float}
\usepackage{subfigure}
\usepackage{tabularx}
\usepackage{xcolor}
\usepackage{color}
\usepackage{siunitx}
\usepackage{xr}
\usepackage{amsmath} 
\usepackage{mathtools}
\usepackage[normalem]{ulem}
\usepackage{natbib}
\usepackage[caption=false]{subfig}
\usepackage{afterpage}
\usepackage{enumerate}
\usepackage{comment}
\makeatletter
\newcommand*{\rom}[1]{\expandafter\@slowromancap\romannumeral #1@}
\makeatother

\begin{document}

\newcommand{\vk}{{\vec{k}}}
\newcommand{\lsy}[1]{{\color{red} #1}}
\newcommand{\duc}[1]{{\color{blue} #1}}
\newcommand{\beq}{\begin{equation}}
\newcommand{\eeq}{\end{equation}}
\newcommand{\vS}{\vec{S}}
\newcommand{\an}[1]{{\color{magenta} #1}}

\title{Noncollinear Antiferromagnetic Order and Effect of Spin-Orbit Coupling \\in Spin-1 Honeycomb Lattice}

\author{Shuyi Li}
\affiliation{Department of Physics and Astronomy, Rice University, Houston, TX 77005, USA}

\author{Manh Duc Le}
\affiliation{ISIS Neutron and Muon Source, Rutherford Appleton Laboratory, Chilton, Didcot, OX11 0QX, UK}

\author{Vaideesh Loganathan}
\affiliation{Department of Physics and Astronomy, Rice University, Houston, TX 77005, USA}

\author{Andriy H. Nevidomskyy}
\email[Correspondence e-mail address: ]{nevidomskyy@rice.edu}
\affiliation{Department of Physics and Astronomy, Rice University, Houston, TX 77005, USA}

\date{\today}

\begin{abstract}
Motivated by the recently synthesized insulating nickelate Ni$_2$Mo$_3$O$_8$, which has been reported to have an unusual non-collinear magnetic order of Ni$^{2+}$ $S=1$ moments  with a nontrivial angle between adjacent spins, we construct an effective spin-1 model on the honeycomb lattice, with the exchange parameters determined with the help of first principles electronic structure calculations. The resulting bilinear-biquadratic model, supplemented with the realistic crystal-field induced anisotropy, favors the collinear N\'eel state.
We find that the crucial key to explaining the observed noncollinear spin structure is the inclusion of the  Dzyaloshinskii--Moriya (DM) interaction between the neighboring spins. By performing the variational mean-field and linear spin-wave theory (LSWT) calculations, we determine that a realistic value of the DM interaction $D\approx 2.78$~meV is sufficient to quantitatively explain the observed angle between the neighboring spins. We furthermore compute the spectrum of magnetic excitations within the LSWT and random-phase approximation (RPA) which should be compared to future inelastic neutron measurements.
\end{abstract}


\maketitle

\section{Introduction}
Recent experimental and theoretical advances in frustrated magnetism, in particular the realization of the Kitaev--Heisenberg model~\cite{kitaev2006,jackeli2009} in the honeycomb lattice materials Na$_2$IrO$_3$~\cite{singh-Na213_2010}, Li$_2$IrO$_3$~\cite{singh-A213_2012}, $\alpha$-RuCl$_3$~\cite{plumb2014}, and H$_3$LiIr$_2$O$_6$~\cite{kitagawa-H3LiIr2O6_2018} 
have sparked much interest in the study of quantum magnets with the honeycomb lattice structure. 
%
%
%
By comparison, honeycomb materials with spin-1 moments have received relatively little attention. Arguably, a larger value of spin makes it more amenable to a semi-classical description, although quantum effects are undeniably important to understand, for instance, the gapped nature of the Haldane ground state in spin-1 chains~\cite{haldane1983,affleck1989}. At the same time, the effect of orbital degrees of freedom and spin-orbit interactions can lead to complex phenomena and a lack of long-range magnetic ordering in spin-1 materials, such as in a recently reported diamond-lattice system NiRh$_2$O$_4$~\cite{chamorro-diamond2018}. In the case of honeycomb spin-1 materials, the same mechanism that was identified as a source of compass-like Kitaev interactions~\cite{jackeli2009} can result in potentially rich physics, including perhaps spin-liquid ground states. In this paper, we set ourselves a less ambitious task and focus on elucidating the puzzling nature of the noncollinear ground state reported recently in a layered honeycomb lattice oxide Ni$_2$Mo$_3$O$_8$~\cite{nafexp}, as shown in Fig.~\ref{fig_NAF3D}. While specific to this material, the present work has wider ramifications for the interplay of frustrations and spin-orbit coupling in spin-1 systems. 

Most of the known spin-1 honeycomb lattice materials are comprised of Ni$^{2+}$ ions, with the strong Hund's coupling leading to spin \mbox{$S=1$} on each site. Similar to spin-$1/2$ case, the vast majority of honeycomb lattice materials, such as $A_3$Ni$_2$SbO$_6$ ($A=\,$Li, Na)~\cite{exp1-zigzag}, Na$_3$Ni$_2$BiO$_6$~\cite{exp2-zigzag} and Li$_3$Ni$_2$BiO$_6$~\cite{exp3}  order in the zigzag pattern depicted in Fig.~\ref{Fig_2d}(f). The N\'eel order shown in Fig.~\ref{Fig_2d}(d) is also possible, as realized for instance in BaNi$_2$V$_2$O$_8$~\cite{BaNiVO-Neel}, while the stripe order is very rare, so far only observed in Ba$_2$Ni(PO$_4$)$_2$ where it is argued to be due to a strong inter-layer exchange coupling~\cite{exp4-stripe,exp4-erratum}. In all the aforementioned cases, the reported magnetic order is collinear, in stark contrast to the material studied here, Ni$_2$Mo$_3$O$_8$, which was reported~\cite{nafexp} to have a noncollinear magnetic structure depicted schematically in Fig.~\ref{fig_NAF3D}. It is also qualitatively different from other molybdenum oxides with the same hexagonal space group such as Fe$_2$Mo$_3$O$_8$ and Mn$_2$Mo$_3$O$_8$, which realize either a collinear antiferromagnetic
or a ferrimagnetic state~\cite{A238_1,A238_2,A238_3}.

In this work, we show that the key to understanding the noncollinear nature of the magnetic ordering in Ni$_2$Mo$_3$O$_8$ is the Dzyaloshinskii--Moriya (DM) interaction that arises due to spin-orbit coupling~\cite{dm1,dm2,dm3,Multiferroics}. From the symmetry analysis, the DM vectors are uniquely determinded by Moriya rules\cite{dm2}. In combination with the exchange couplings computed from first principles density functional theory (DFT), this allows us to reproduce the experimentally reported magnetic structure. We further compute the generalized phase diagram, with the angle between the two neighboring spins being a function of the DM interaction strength and exchange parameters of the model. Importantly, inclusion of the biquadratic spin-spin interactions of the type $(\vS_i\cdot \vS_j)^2$ is necessary to both fit the \textit{ab initio} results and predict the correct noncollinear magnetic structure.

The remainder of this article is organized as follows. We present an effective spin-1 model in section~\ref{sec:model}. Various competing spin configurations and their mean field energies are introduced in section~\ref{sec:spin_conf} and \ref{sec:MFT}, followed by the details of determination of spin exchange couplings from \textit{ab initio} calculations in section~\ref{sec:dft}. We analyze the single spin anisotropy term from crystal field theory in section~\ref{sec:SIA}. In section~\ref{sec:results}, we compute the phase diagram of the model with and without Dzyaloshinskii--Moriya interactions, demonstrating that the latter are crucial to reproduce the experimentally reported noncollinear magnetic state. Then, in section~\ref{sec:LSW}, we perform linear spin wave theory calculations in competing phases to capture the quantum fluctuations around mean-field solutions. Finally, we conclude with the discussion and outlook in section~\ref{sec:discussion}.

\section{Model}\label{sec:model}

Ni$_2$Mo$_3$O$_8$ crystallizes in the layered structure characterized by the non-centrosymmetric hexagonal space group $P6_3mc$~\cite{nafexp}, with the Ni$^{2+}$ magnetic ions forming a buckled hexagonal structure in each layer shown schematically in Fig.~\ref{fig_NAF3D}.
%
There are two inequivalent Ni sites in this bipartite structure, with alternating atoms having either octahedral or tetrahedral coordination by oxygen ions. The magnetic moments on these two sublattices form two interpenetrating triangular lattices, with an angle $\alpha$ with each other, as depicted in Fig.~\ref{fig_NAF3D}, resulting in a noncollinear antiferromagnetic (NCAF) order. 


In order to model the spin 1 moments on Ni$^{2+}$ ($3d^8$) ions in this material, we adopt a bilinear-biquadratic spin-1 quantum Heisenberg model, at first without taking spin-orbit coupling into account:
\begin{equation}
\label{HwithoutDM}
\begin{aligned}
\mathcal{H}_\text{eff} &= 
\sum_{\langle ij\rangle}J_{1} \vec{S}_{i}\cdot\vec{S}_{j} + K_{1} (\vec{S}_{i}\cdot\vec{S}_{j})^2 \\
&+\sum_{\langle\langle ij\rangle\rangle,T}J_{2T} \vec{S}_{i}\cdot\vec{S}_{j} + K_{2T} (\vec{S}_{i}\cdot\vec{S}_{j})^2\\
&+\sum_{\langle\langle ij\rangle\rangle,O}J_{2O} \vec{S}_{i}\cdot\vec{S}_{j} + K_{2O} (\vec{S}_{i}\cdot\vec{S}_{j})^2\\
&+\mathcal{H}_\text{A}.
\end{aligned}
\end{equation} 
where $J_1, K_1$ are the nearest-neighbour Heisenberg and biquadratic couplings, while $J_{2T}, K_{2T}$ ($J_{2O}, K_{2O}$) describe the second-neighbor spin-spin interactions between tetrahedral (octahedral) sites, respectively. As we shall show below in section~\ref{sec:dft}, the inclusion of biquadratic spin-spin interactions is crucial to correctly reproduce the magnon excitation spectrum and in order to match the energy differences between the various magnetically ordered reference states obtained from \textit{ab initio} calculations. 
We also take into account the different crystal-field effects on the tetrahedrally and octahedrally coordinated Ni ions, which results in the effective single-ion spin anisotropy Hamiltonian $\mathcal{H}_\text{A}$. We shall discuss the form of $\mathcal{H}_\text{A}$ in section~\ref{sec:SIA} below and in Appendix~\ref{app:cef}.\\


As advertised earlier, the inclusion of the spin-orbit coupling in the form of the Dzyaloshinskii--Moriya (DM) interactions among the spins is essential to reproduce the noncollinear magnetic structure. This will be discussed in detail in section~\ref{sec:results}, here we write down the DM Hamiltonian for completeness: 
\begin{equation}
\label{dmh}
\mathcal{H}_{\text{DM}}=\frac{1}{2}\sum_{ij}\vec{D}_{ij}\cdot (\vec{S}_i\times\vec{S}_j),
\end{equation} 
where $\vec{D}_{ij}$ is a vector whose direction can be determined by Moriya's rules\cite{dm2}, to be discussed in section~\ref{sec:results}.

\section{Single-ion Crystal field Analysis} \label{sec:SIA}
Due to the interplay of spin-orbit coupling and crystal field effects on Ni ion, there are single-ion spin anisotropy terms in the Hamiltonian. Because of the three-fold rotation symmetry $C_3$ in the $P6_3mc$ space group, the crystal field Hamiltonian under Wybourne normalization is given by (see Appendix~\ref{app:cef2} for more details)
\begin{equation}
\label{eq:CFE}
    \mathcal{H}_{cf}= L_{20}\theta_2\hat{T}_{20} + L_{40}\theta_4\hat{T}_{40} + L_{43}\theta_4\hat{T}_{43},
\end{equation}
where $L_{lm}$ are the crystal field parameters, $\theta_l$ are the Stevens factors, and $\hat{T}_{lm}$ are tensorial Stevens--Wybourne operators. We use the point charge model as discussed in Appendix~\ref{app:cef} which yields the following crystal field parameters:
\begin{equation}
\label{eq:CFT}
\begin{aligned}
    L_{20}^T& = +626 \,\text{meV}, L_{40}^T = +307 \, \text{meV}, L_{43}^T = +764 \, \text{meV}\\
    L_{20}^O& = -166 \,\text{meV}, L_{40}^O = -1392 \, \text{meV}, L_{43}^O = 1623 \, \text{meV}.
\end{aligned}
\end{equation}

While the Hamiltonian~\eqref{eq:CFE} acts on the components of the orbital momentum, the spin degrees of freedom feel the effect of anisotropies because of the spin-orbit coupling
$\lambda\vec{S}\cdot\vec{L}$. Estimating the coupling constant $\lambda\approx-40$~meV~\cite{soc}, as is typical for Ni ions, we treat it as a perturbation. While the first order of the perturbation vanishes identically, the correction to the energy in the second order perturbation theory is of the form
\begin{equation} \label{eq:2ndorderCFT}
    E_{\text{2nd}}=\lambda^2\sum_{i,j}\Lambda_{ij}S^iS^j.
\end{equation}
This results in the effective single-ion spin anisotropy Hamiltonian
\begin{equation}
\label{eq:anisotropy}
    \mathcal{H}_\text{A}=\sum_{T}\gamma_{T}(S_i^z)^2+\sum_{O}\gamma_{O}(S_i^z)^2
\end{equation}
with $\gamma_{T}=30.41$ meV, $\gamma_{O}=-0.53$ meV (see Appendix~\ref{app:cef2} for details of the derivation).
These values imply the tetrahedral sites strongly prefer to lie in the $xy$-plane, whilst the octahedral sites a have weak preference to align along $z$.

\section{Spin configurations}
\label{sec:spin_conf}

\begin{figure}[tb]
    \includegraphics[width=0.35\textwidth]{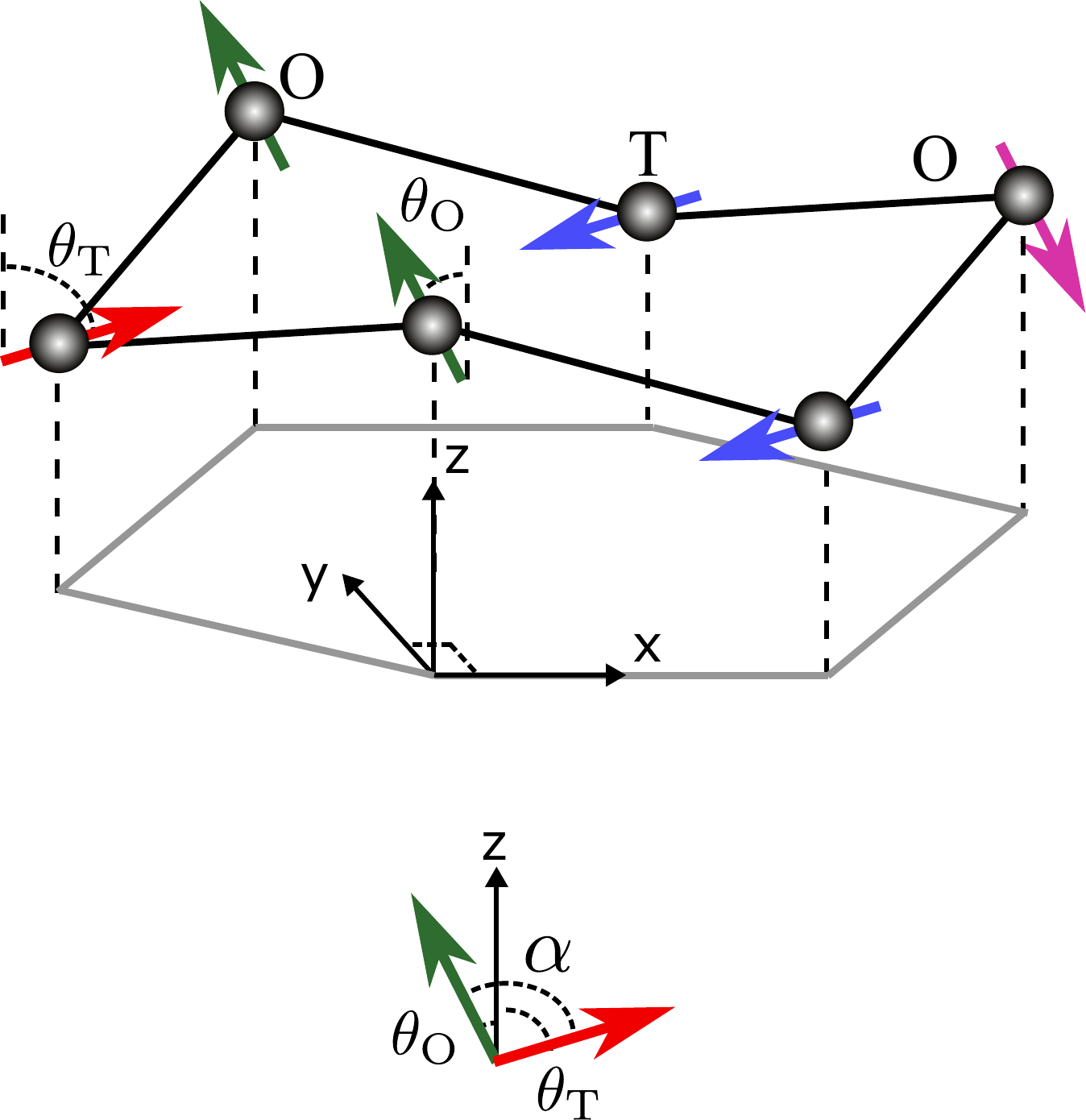}
    \caption{Depiction of the NCAF state in Ni$_2$Mo$_3$O$_8$~\cite{nafexp}. The Ni$^{2+}$ ions are represented by grey spheres. The coordinate system is set up by projecting from the non-centrosymmetric buckled honeycomb lattice onto a regular hexagon in $ab$-plane (marked by light grey color). The two inequivalent Ni sites have either a tetrahedral (T) or octahedral (O) oxygen coordination. On each of these two sublattices, the spins form a collinear antiferromagnetic order with magnetic ordering wave vector $\vec{q}_m=(\frac{2\pi}{3a},0)$, where $a$ is the side length of the projected regular hexagon in $xy$-plane. When moments lie in the $xz$-plane, as is the case in Ni$_2$Mo$_3$O$_8$~\cite{nafexp},  
    the angle between the nearby T and O sites is $\alpha=\theta_T+\theta_O$, as depicted in the lower panel.}
    \label{fig_NAF3D}
\end{figure}

In Ni$_2$Mo$_3$O$_8$, the state we are mainly interested in is the NCAF state  shown in Fig.~\ref{fig_NAF3D}. The neutron scattering analysis~\cite{nafexp} shows that the moments form a coplanar structure in the $xz$-plane, with $x$ axis pointing along one of the hexagonal bonds and the $z$ axis being the hexagonal $c$-axis of the crystal, as indicated in Fig.~\ref{fig_NAF3D}.
In the honeycomb lattice, tetrahedral (T) and octahedral (O) sites form two interpenetrating triangular sublattices. In order to completely characterize various spin states, we introduce the polar angles $\theta_T$ and $\theta_O$ relative to the $z$-axis on each sublattice, and the asimuthal angles $\phi_T$ and $\phi_O$ with the $x$-axis, respectively. 
The angle $\alpha$ between the neighboring spins on the two sublattices is then given by  
\begin{equation}
\cos\alpha=\sin\theta_T\sin\theta_O\cos(\phi_T-\phi_O)+\cos\theta_T\cos\theta_O.    \label{eq:alpha}
\end{equation}
Since the moments in the experimental NCAF phase lie in the $xz$ plane, the asimuthal angles are either $0$ or $\pi$, and moreover $|\phi_T-\phi_O|=\pi$. We shall assume this to be the case in the following. From Eq.~(\ref{eq:alpha}), it then follows that the angle $\alpha$ between the two spins is 
\begin{equation}
\alpha=\theta_T+\theta_O, \label{eq:alpha2}
\end{equation}
as depicted in the bottom of Fig.~\ref{fig_NAF3D}.
For convenience, if $\alpha>180^{\circ}$, it is equivalent to use $\alpha'=360^{\circ}-\alpha$ as the angle between two spin directions. Thus it is sufficient to only consider $0^{\circ}\leq \alpha\leq 180^{\circ}$.

\begin{figure}[h]
\includegraphics[width=0.4\textwidth]{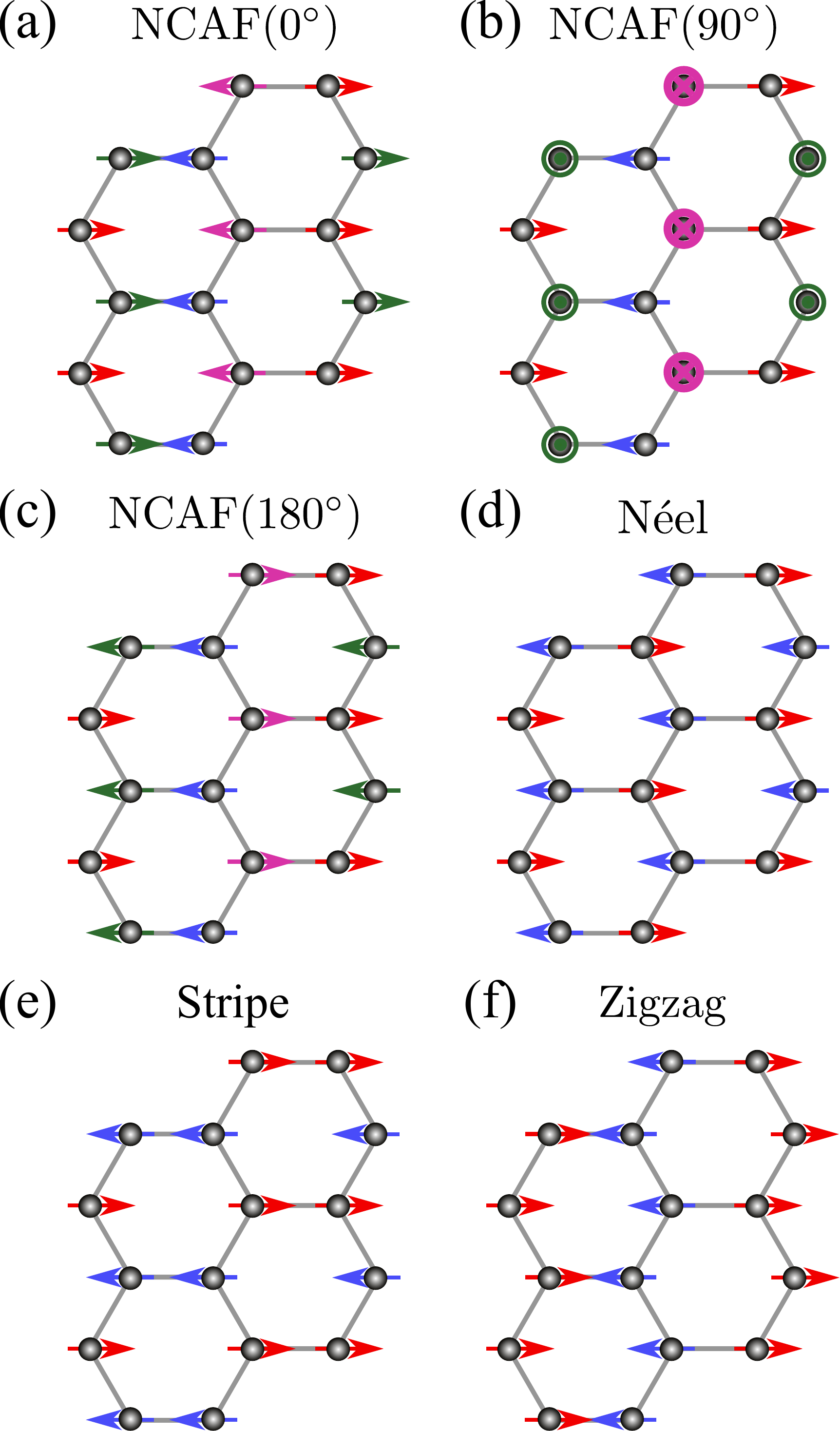}
\caption{The depiction of the noncollinear antiferromagnet configuration with (a) $\alpha=0$, (b)~$\alpha=90^{\circ}$ and (c) $\alpha=180^{\circ}$. Also shown are typical collinear configurations on the honeycomb lattice: (d)~N\'eel state, (e) stripe state, and (f) zigzag state.}
\label{Fig_2d}
\end{figure}

While the experimental ground state of Ni$_2$Mo$_3$O$_8$ is noncollinear,
it is instructive to look at the various collinear magnetic orders obtained by setting $\alpha=0$ or $\alpha=180^{\circ}$, depicted in Fig.~\ref{Fig_2d} (a) and (c), respectively. 
In the honeycomb lattice model, one often considers three important collinear spin ordered states: N\'eel, stripe and zigzag states, depicted in Fig.~\ref{Fig_2d} (d), (e) and (f). As the figure illustrates, the zigzag and stripe orderx correspond to $\alpha=0$ and $\alpha=180^{\circ}$, respectively, and one can think of a noncollinear NCAF states with generic value of $\alpha$ as lying in-between these two limiting cases, such as for instance the special case with $\alpha=90^{\circ}$ depicted in Fig.~\ref{Fig_2d}(b).

\begin{table}
\caption{Spin configurations and magnetic moments of the two experimental fits to the neutron scattering data on Ni$_2$Mo$_3$O$_8$, inferred from Ref.~\onlinecite{nafexp}. The angles $\theta$ and $\phi$ for the two sublattices are defined in the text.}
\begin{tabular}{|c||c|c|c|c|c|c||c|c|}   
\hline 
  {}& $\theta_T$ & $\phi_T$ & $ M_{T}$ & $\theta_O$ & $\phi_O$ & $M_{O}$ & $\alpha$ & $\alpha'$\\
\hline      
  Fit 1  & 85$^{\circ}$ & 180$^{\circ}$ & 1.727$\mu_{B}$ & 145$^{\circ}$ & 0$^{\circ}$ & 1.431$\mu_{B}$ & {\bf 230$^{\circ}$} & {\bf 130$^{\circ}$}\\
\hline 
  Fit 2  & 124$^{\circ}$ & 180$^{\circ}$ & 1.997$\mu_{B}$ & 87$^{\circ}$ & 0$^{\circ}$ & 0.891$\mu_{B}$ & {\bf 211$^{\circ}$} & {\bf 149$^{\circ}$}\\
\hline
\end{tabular}
\label{table1}
\end{table}

We note that the experimental study in Ref.~\onlinecite{nafexp} reports two possible magnetic structures, with different sizes and directions of the magnetic moments; which we summarize in Table~\ref{table1}. Both structures provide an equally good fit to the  neutron scattering refinements, however as we shall show below, our theoretical analysis suggests that the experimental structure 1, with $\alpha'=130^{\circ}$, is most likely realized in Ni$_2$Mo$_3$O$_8$.



\section{Mean field energy Of different magnetic orders}\label{sec:MFT}

Our goal is to obtain accurate estimates of the exchange couplings in the model Hamiltonian Eq.~(\ref{HwithoutDM}) from first principles calculations. To do this, we first evaluate analytically the mean field (MF) energies of several reference ordered states $|\psi\rangle$, namely a ferromagnet (FM), N\'eel, stripe and zigzag states, by using spin-1 product states as a MF ansatz:
\begin{equation}
\label{eq:productstate}
    |\psi\rangle=\prod_i\otimes|\vec{S}_i\rangle,
\end{equation}
where $|\vec{S}_i\rangle$ on a given site is the maximum-weight eigenstate along the local direction $(\theta_i,\phi_i)$ given by
\begin{equation}
\label{eq:spinstate}
\begin{aligned}
    |\vec{S}_i\rangle&=e^{-i\phi_i}\cos^2{\frac{\theta_i}{2}}~|1\rangle+e^{i\phi_i}\sin^2{\frac{\theta_i}{2}}~|-1\rangle\\
    &+\sqrt{2}\sin{\frac{\theta_i}{2}}\cos{\frac{\theta_i}{2}}~|0\rangle.
\end{aligned}
\end{equation}
By choosing different local $(\theta_i,\phi_i)$ directions, we can capture different ordered states. For instance, the ferromagnetic state is given by $[\theta_T=\theta_O,\phi_T=\phi_O]$, whereas the 
 N\'eel state is accommodated by  $[\theta_T=\theta_O+180^{\circ},\phi_T=\phi_O]$. The resulting mean-field expressions for the energies of various reference states are as follows:
\begin{equation}
\label{ewodm}
\begin{aligned}
\mathcal{E}_{\text{FM}}=&\frac{3}{2}J_1+\frac{3}{2}J_{2T}+\frac{3}{2}J_{2O}\\
+&\frac{\gamma_T}{4}(\cos^2\theta_T+1)+\frac{\gamma_O}{4}(\cos^2\theta_O+1),\\
\mathcal{E}_{\text{N\'eel}}=&-\frac{3}{2}J_1+\frac{3}{2}K_1+\frac{3}{2}J_{2T}+\frac{3}{2}J_{2O}\\
+&\frac{\gamma_T}{4}(\cos^2\theta_T+1)+\frac{\gamma_O}{4}(\cos^2\theta_O+1),\\
\mathcal{E}_{\text{Stripe}}=&-\frac{1}{2}J_1+K_1-\frac{1}{2}J_{2T}+K_{2T}-\frac{1}{2}J_{2T}+K_{2O}\\
+&\frac{\gamma_T}{4}(\cos^2\theta_T+1)+\frac{\gamma_O}{4}(\cos^2\theta_O+1),\\
\mathcal{E}_{\text{Zigzag}}=&\frac{1}{2}J_1+\frac{1}{2}K_1-\frac{1}{2}J_{2T}+K_{2T}-\frac{1}{2}J_{2T}+K_{2O}\\
+&\frac{\gamma_T}{4}(\cos^2\theta_T+1)+\frac{\gamma_O}{4}(\cos^2\theta_O+1).\\
\end{aligned}
\end{equation}

Because the equations (\ref{ewodm}) are linearly dependent, we introduce more reference states in order to be able to determine the exchange couplings uniquely (see Appendix~\ref{app:MF} for more details). 
For future reference, we provide here the MF expression for the energy of the NCAF state for an arbitrary angle $\alpha$ between the spins on T and O sites, as defined in Fig.~\ref{fig_NAF3D} and in Eq.~(\ref{eq:alpha}): 
\begin{equation}
\label{ncaf}
\begin{aligned}
\mathcal{E}_{\text{NCAF}}=&\frac{1}{2}J_1\cos\alpha+K_1(\frac{3}{8}\cos^2\alpha-\frac{1}{4}\cos\alpha+\frac{3}{8})\\
-&\frac{1}{2}J_{2T}+K_{2T}-\frac{1}{2}J_{2T}+K_{2O}\\
+&\frac{\gamma_T}{4}(\cos^2\theta_T+1)+\frac{\gamma_O}{4}(\cos^2\theta_O+1).
\end{aligned}
\end{equation}


\section{DFT analysis}\label{sec:dft}

\begin{figure}[t]
    \centering
    \includegraphics[width=0.42\textwidth]{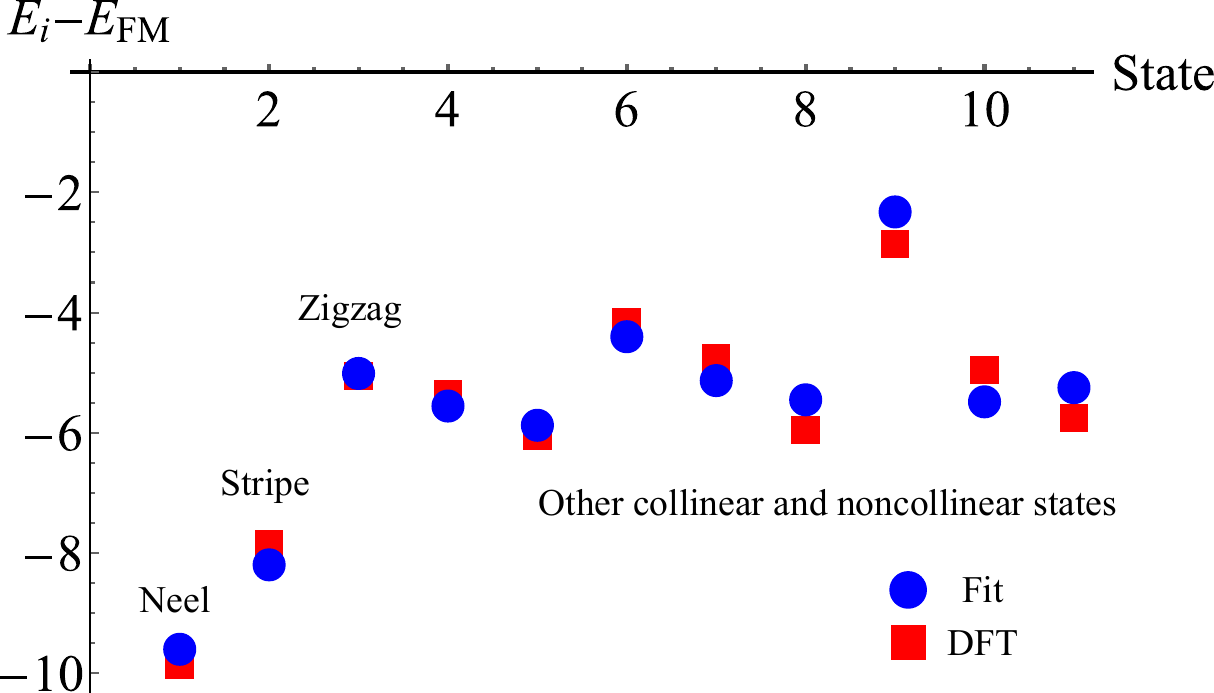}
    \caption{Least squares fitting of
the mean-field energies for the model in Eq.~\eqref{HwithoutDM}  to  the \emph{ab initio} DFT energies for 11 reference magnetically ordered states (relative to the FM state energy). The red points indicate the \emph{ab initio} energies, while the blue points are the MF energy difference with the best fitting parameters quoted in Eq.~\eqref{eq:DFT}. 
    }
    \label{fig:fit}
\end{figure}

We have performed \emph{ab initio} density functional theory calculations on Ni$_2$Mo$_3$O$_8$ (see Appendix~\ref{app:DFT} for details) in various spin-ordered states, both collinear and noncollinear, and computed the corresponding energies. 
By fixing a collinear ordered state and choosing different \textit{global} rotations, we have found that DFT captures the single-ion anisotropy poorly -- the resulting energy differences are about 0.1~meV per site, much less than expected from the relatively large value of $\gamma_T$ in Eq.~\eqref{eq:anisotropy}. This is a known effect to do with the inaccuracy of capturing crystal-field splittings and the approximate way in which the spin-orbit coupling is treated in typical \emph{ab initio} codes.
Similarly, the Dzyaloshinskii--Moriya interactions are not captured well at the level of DFT. Thus, we used the DFT reference energies to determine only six unknown parameters in Eq.~\eqref{HwithoutDM}: $J_1, J_{2T}, J_{2O}, K_1, K_{2T}$ and $K_{2O}$. 
Substituting these \textit{ab initio} energies into the left-hand side of the mean-field expressions in Eq.~(\ref{ewodm}) and other reference states (see Appendix~\ref{app:MF}), we use the least-square fitting to determine the set of the exchange coefficients. 
In total, 12 reference states, and hence 11 energy differences have been used, resulting in the excellent quality of the least-square fit ($R^2=0.956$) shown in Fig.~\ref{fig:fit}.  
The obtained values of the fitting parameters are as follows:
\begin{equation}
\label{eq:DFT}
\begin{aligned}
    J_1& = +2.62 \,\text{meV}, &K_1 = -1.13 \, \text{meV},\\
    J_{2T}& = +0.35 \, \text{meV}, &K_{2T} = -0.35\, \text{meV},\\
    J_{2O}& = +0.41 \, \text{meV}, &K_{2O} = +0.09\, \text{meV}.
\end{aligned}
\end{equation}

The most important conclusion for this work is that the value of $K_1$ is negative and non-negligible compared to the Heisenberg exchange $J_1$. The presence of such biquadratic term in the model Eq.~(\ref{HwithoutDM}) is important to correctly capture the physics of spin 1 interactions, as was proven to be the case in other $3d$ metals with spin-1 moments, notably the iron pnictides and chalcogenides. There, one also finds negative and relatively large values of $K_1$ from first principles caculations~\cite{Wysocki2011,Glasbrenner2015}, and it turns out to be essential to correctly describe the magnon dispersion in inelastic neutron scattering~\cite{Harriger2011,Yu2012,Ergueta2015,Ergueta2017}. In the present case, we shall show that the presence of $K_1$ term affects the relative stability of the N\'eel and noncollinear magnetic states (see section~\ref{sec:DM}).

The \textit{ab initio} electronic structure calculations reveal additional information about the magnetic properties of Ni$_2$Mo$_3$O$_8$. The magnitude of the magnetic moment remains unchanged across the various ordered states and is dominated by the Hund's coupled spin contribution of $\langle \hat{M_S}\rangle = 1.45$ $\mu_B$ per Ni for both types (T,O) of Ni atoms. There is also an orbital moment contribution, which is an order of magnitude smaller, $\langle \hat{M}_L^{(T)}\rangle = 0.18\,\mu_B$ on the tetrahedral Ni ion and $\langle\hat{M}_L^{(O)}\rangle = 0.12\,\mu_B$ on the octahedral ion (the slightly different values are due to the difference in the crystal field environment on the two sites). 
The total magnetic moment is thus predicted to be $\langle \hat{M}_J\rangle = \langle \hat{M}_S\rangle + \langle \hat{M}_L\rangle = 1.63\,\mu_B$ on the tetrahedral site and $1.57\,\mu_B$ on the octahedral site. These values of the moments are closer to the first of the two experimental fits from Ref.~\onlinecite{nafexp} shown in Table~\ref{table1}. 


\section{Results} \label{sec:results}
Having estimated the spin exchange couplings from the \textit{ab initio} calculations and the single-ion anisotropy from point charge model (see section~\ref{sec:SIA}), we now proceed to compute the theoretical phase diagram as a function of these parameters, in two regimes: first without the Dzyaloshinskii--Moriya interaction using the effective spin model in Eq.~(\ref{HwithoutDM}), and then incorporating it into the model. As we shall demonstrate, the DM interaction is crucial in order to correctly capture the non-collinear antiferromagnetic state observed experimentally~\cite{nafexp} in Ni$_2$Mo$_3$O$_8$.

\subsection{Results without DM interaction}\label{sec:noDM}

\begin{figure}[b]
    \hspace{-0.4cm}
    \includegraphics[width=0.45\textwidth]{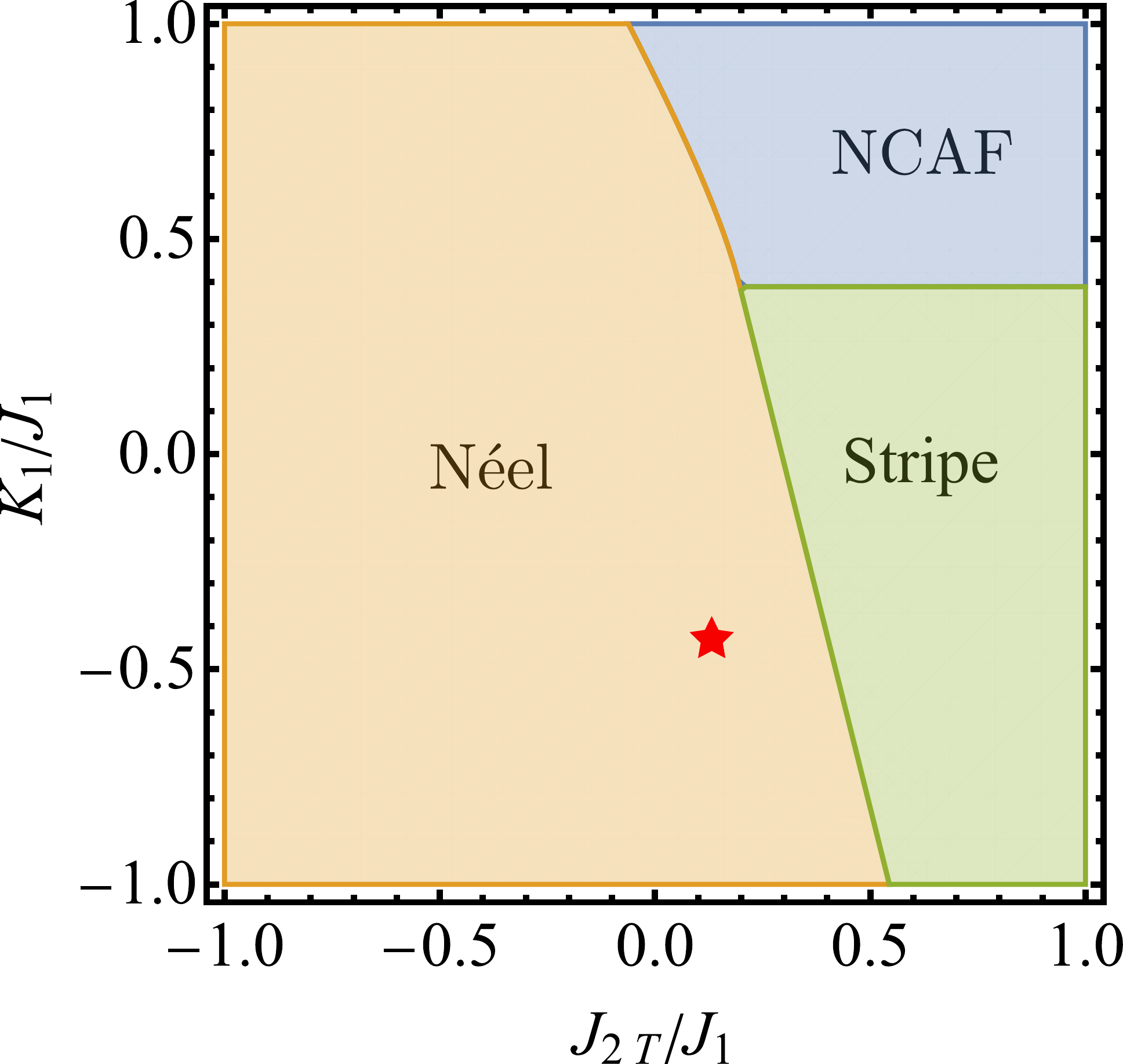}
    \caption{The phase diagram of the model in Eq.~(\ref{HwithoutDM}) in the $({J_{2T}}-{K_1})$ parameter space. The values $J_1=2.62$ meV $J_{2O}=0.41$ meV, $K_{2T}=-0.35$ and $K_{2O}=0.09$meV are fixed as determined from \textit{ab initio} calculations. The red asterisk corresponds to the set of parameters ($J_{2T}=0.35$ meV, $K_1=-1.13$ meV) determined from \textit{ab initio} calculations in Eq.~(\ref{eq:DFT}).}
    \label{fig_phasediagram1}
\end{figure}

Because $\gamma_T$ is positive and large in Eq.~\eqref{eq:anisotropy}, the moments on tetrahedral sites prefer to lie in the $xy$ plane, consistent with the polar angle $\theta_T$ close to 9$0^\circ$ in the experimental Fit 1 in Table~\ref{table1}. For the model parameters in Eq.~(\ref{eq:DFT}), the classical Luttinger-Tisza method shows that the ground state has N\'eel order (see Appendix \ref{app:LT} for more details). This conclusion is corroborated by the mean-field calculations -- indeed, by comparing the expected energies of the different magnetic states in Eq.~(\ref{ewodm}) and Eq.~(\ref{ncaf}), we find that in the absence of the DM interaction, the collinear N\'eel phase dominates a large portion of the phase diagram, with both the T and O moments lying in the $xy$ plane.
This is illustrated in Fig.~\ref{fig_phasediagram1} (the parameters in Eq.~(\ref{eq:DFT}) are shown with an asterisk, which lies inside the N\'eel phase), where we have fixed the values of $J_1$, $J_{2O}$, $K_{2T}$, $K_{2O}$, $\gamma_T$ and $\gamma_O$, and show the phase diagram as a function of the relative strength of $J_{2T}$ and $K_1$. We have set $J_1>0$ since both the experiment and our \textit{ab initio} calculations indicate that the nearest-neighbor exchange is antiferromagnetic, see Eq.~\eqref{eq:DFT}.

Competing with the N\'eel state is the noncollinear state parametrized by the relative angle  $\alpha=\theta_T + \theta_O$ on the two sublattices (see Eq.~\eqref{eq:alpha} and Fig.~\ref{fig_NAF3D} for the meaning of $\alpha$). Note that the mean-field energy of such a non-collinear state $\mathcal{E}_{NCAF}$ in Eq.~(\ref{ncaf}) is a function of $\theta_T$ and the relative angle $\alpha$.
We determine the optimal angles $\alpha$ and $\theta_T$
by minimizing the energy $\frac{\partial \mathcal{E}_{NCAF}}{\partial \alpha}=0$ and $\frac{\partial \mathcal{E}_{NCAF}}{\partial \theta_T}=0$, which results in a set of coupled equations
\begin{equation}
\label{eq:alpha-opt}
\begin{aligned}
\gamma_T\sin2\theta_T&=(-2J_1-3K_1\cos\alpha+K_1)\sin\alpha,\\
\gamma_T\sin2\theta_T&=\gamma_O\sin2(\alpha-\theta_T).
\end{aligned}
\end{equation}
Note that the stripe phase shown in Fig.~\ref{Fig_2d}(e) is a special case of the NCAF state with $\alpha=\pi$, and we find the stripe state to be stabilized for sufficiently large $J_{2T}$, provided $K_1$ is below $\lesssim 0.4 J_1$, as shown in Fig.~\ref{fig_phasediagram1}. 

Our \textit{ab initio} calculations indicate that $K_1$ is negative, and the set of exchange parameters computed from DFT (shown with an asterisk in Fig.~\ref{fig_phasediagram1}) lies very close to the boundary between the N\'eel and the stripe phase. It is clear from Figure~\ref{fig_phasediagram1} that unless the value of $K_1$ is sufficiently large and positive (namely, $K_1>0.39J_1$), which is not the case in our \textit{ab initio} set of parameters, the noncollinear solution will \textit{not} be realized. The inclusion of quantum fluctuations beyond mean-field theory does not alter this conclusion, as will be demonstrated below in Sec.~\ref{sec:LSW}. 
We therefore turn our attention to the effect of Dzyaloshinskii--Moriya interactions, which as we show below,  qualitatively changes the phase diagram.



\subsection{The effect of DM interaction}\label{sec:DM}

As shown above, the Heisenberg model favors collinear spin-ordering. The non-centrosymmetric crystal structure of Ni$_2$Mo$_3$O$_8$ motivates us to consider Dzyaloshinskii--Moriya interactions arising from spin orbit coupling. While it will not affect the energies of the collinear spin configurations such as N\'eel, stripe or zigzag states, the DM interaction can potentially lower the energy of the noncollinear states.

\begin{figure}[h]
    \centering
    \includegraphics[width=0.2\textwidth]{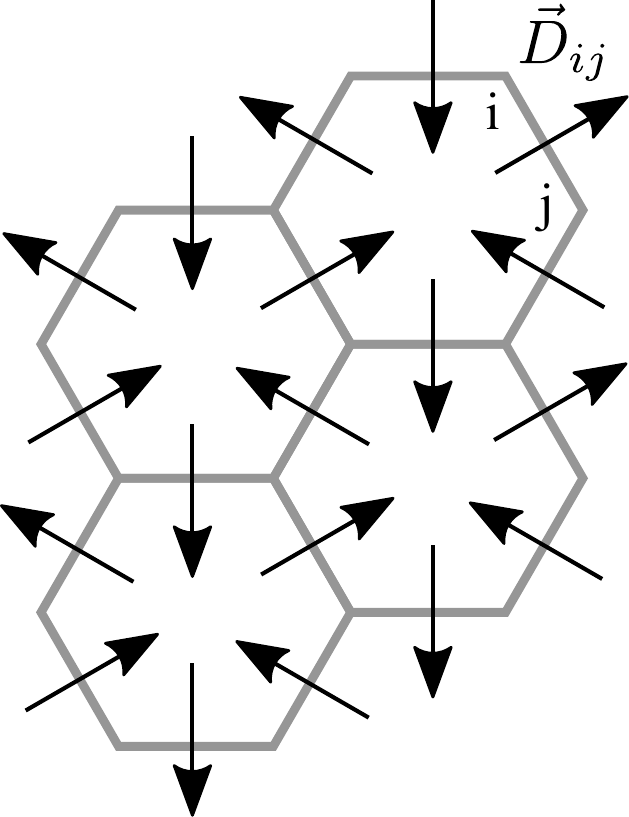}
    \caption{The DM vectors for nearest neighboring bonds of non-centrosymmetric honeycomb lattice in $xy$ plane.}
    \label{fig_DM}
\end{figure}

Consider first the DM interaction between spins on the nearest sites $O$ and $T$. In a non-centrosymmetric honeycomb lattice, there is only one mirror plane that includes both sites, which is perpendicular to the $ab$ plane. From Moriya's rules, the vector $\vec{D}_{OT}$ should be prependicular to this mirror plane, which means that $\vec{D}_{OT}$ lies in the $ab$ plane, and is perpendicular to the bond direction $\vec{OT}$. By the $C_3$ rotational symmetry of the lattice, we can obtain the vectors $\vec{D}_{ij}$ for all nearest neighbor sites, as shown in Fig.~\ref{fig_DM}, which should all have the same magnitude $|\vec{D}_{ij}|=D$. 

\begin{figure}[t!]
    \centering
    \includegraphics[width=0.5\textwidth]{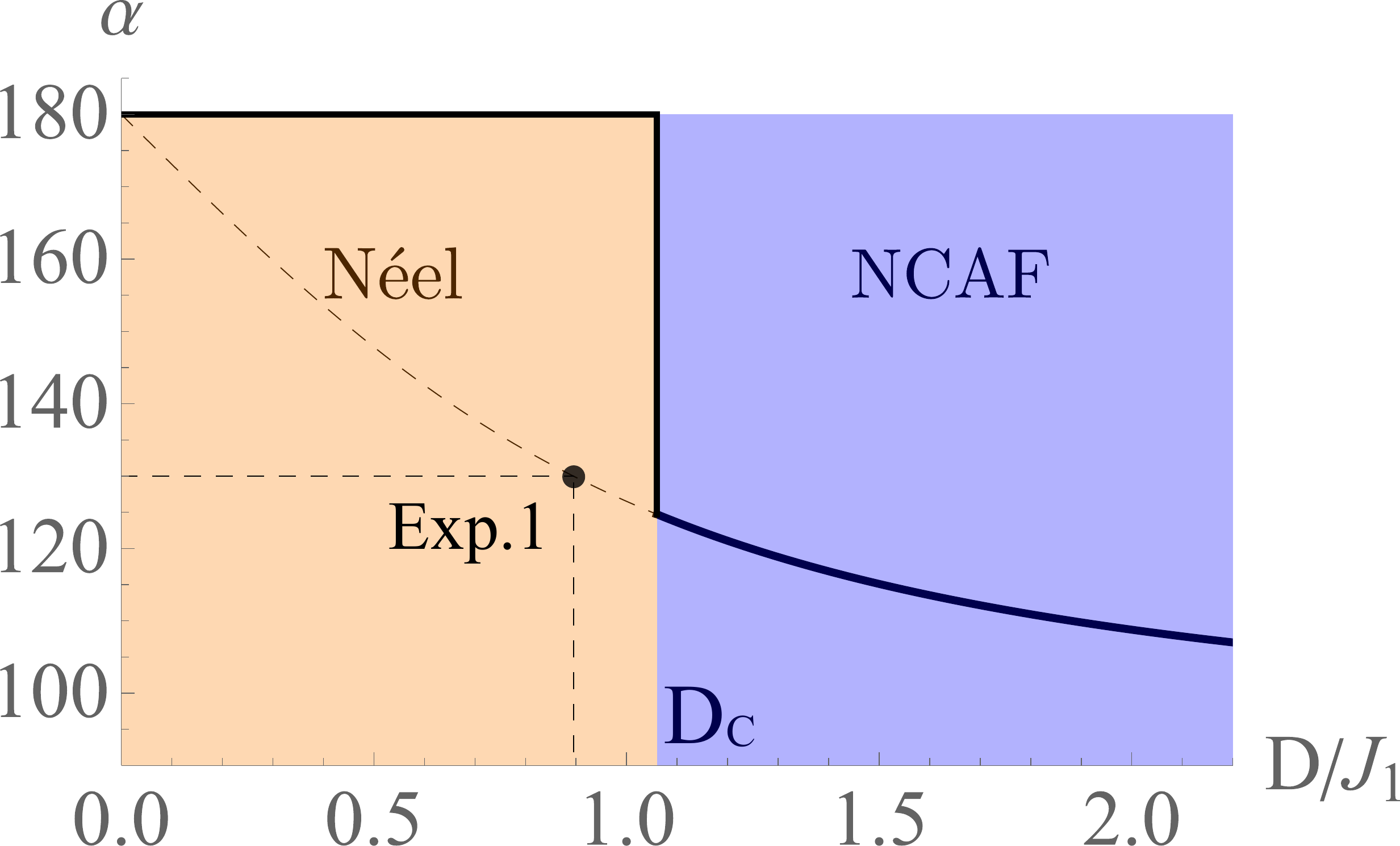}
    \caption{The optimal value of angle $\alpha$ corresponding to the minimum NCAF energy as a function of DM strength $D$, with the exchange parameters fixed at the \textit{ab initio} values in Eq.~(\ref{eq:DFT}). The dashed lines corresponds to the experimental value of $\alpha'=130^{\circ}$, achieved at $D\approx 2.35$ meV.}
    \label{minangle}
\end{figure}

At the expected energy level, the average energy per site of the variational NCAF state is
\begin{equation}
\label{eq:E-DM-NCAF}
\begin{aligned}
\mathcal{E}_{\text{NCAF}}^{'}&=\frac{1}{2}J_1\cos\alpha+K_1(\frac{3}{8}\cos^2\alpha-\frac{1}{4}\cos\alpha+\frac{15}{8})\\
&-\frac{1}{2}J_{2T}+\frac{5}{2}K_{2T}-\frac{1}{2}J_{2O}+\frac{5}{2}K_{2O}\\
&-D(\sin\theta_O\cos\phi_O\cos\theta_T-\sin\theta_T\cos\phi_T\cos\theta_O)\\&+\frac{1}{4}\gamma_T(\cos^2\theta_T+1)+\frac{1}{4}\gamma_O(\cos^2\theta_O+1),
\end{aligned}
\end{equation}
where as before, $\alpha=\theta_T+\theta_O$ is the angle between the spins on the tetrahedral and octahedral sites, as indicated in Fig.~\ref{fig_NAF3D}. Under the assumption that both spins lie in the plane containing the O--T bond, as realized in the experiment ($|\phi_T-\phi_O|=\pi$ in our notation), the Dzyaloshinskii--Moriya term results in the energy contribution
\begin{equation}
\mathcal{E}_{\text{DM}}=D\cos\phi_T\sin\alpha.
\end{equation}
To minimize this energy  we choose, without loss of generality, $\phi_T=180^{\circ}$, which corresponds to the experimental results in Table~\ref{table1}. 
The energy of the NCAF ordered state then becomes
\begin{equation}
\label{eq:EDM}
\begin{aligned}
\mathcal{E}_{\text{NCAF}}^{'}&=\frac{1}{2}J_1\cos\alpha+K_1(\frac{3}{8}\cos^2\alpha-\frac{1}{4}\cos\alpha+\frac{3}{8})\\
&-\frac{1}{2}J_{2T}+K_{2T}-\frac{1}{2}J_{2O}+K_{2O}\\
&-D\sin\alpha\\
&+\frac{1}{4}\gamma_T(\cos^2\theta_T+1)+\frac{1}{4}\gamma_O(\cos^2(\alpha-\theta_T)+1).
\end{aligned}
\end{equation}

By minimizing this energy with respect to the variational parameter $\alpha$ and $\theta_T$, we thus obtain the optimal value of $\alpha$ for a given $D$ (it is clear from Eq.~\ref{eq:EDM} that it suffices to consider $D>0$, since $0\leq \alpha \leq \pi$). The resulting optimal angle as a function of the DM interactions strength $D$ is shown in Fig.~\ref{minangle}. It shows that for $D$ less than a critical value of  $D_c =1.06J_1\approx 2.78$~meV, the N\'eel state is the ground state, and for larger values of the DM interaction, a first order phase transition into the NCAF state takes place, with the angle $\alpha$ jumping to a value $\alpha \lesssim 124.7^\circ$. Note that the critical value of $D_c$ results in the angle close to the experimentally reported $\alpha'=130^\circ$ in Table~\ref{table1}.

\begin{figure}[tb!]
    \centering
    \includegraphics[width=0.45\textwidth]{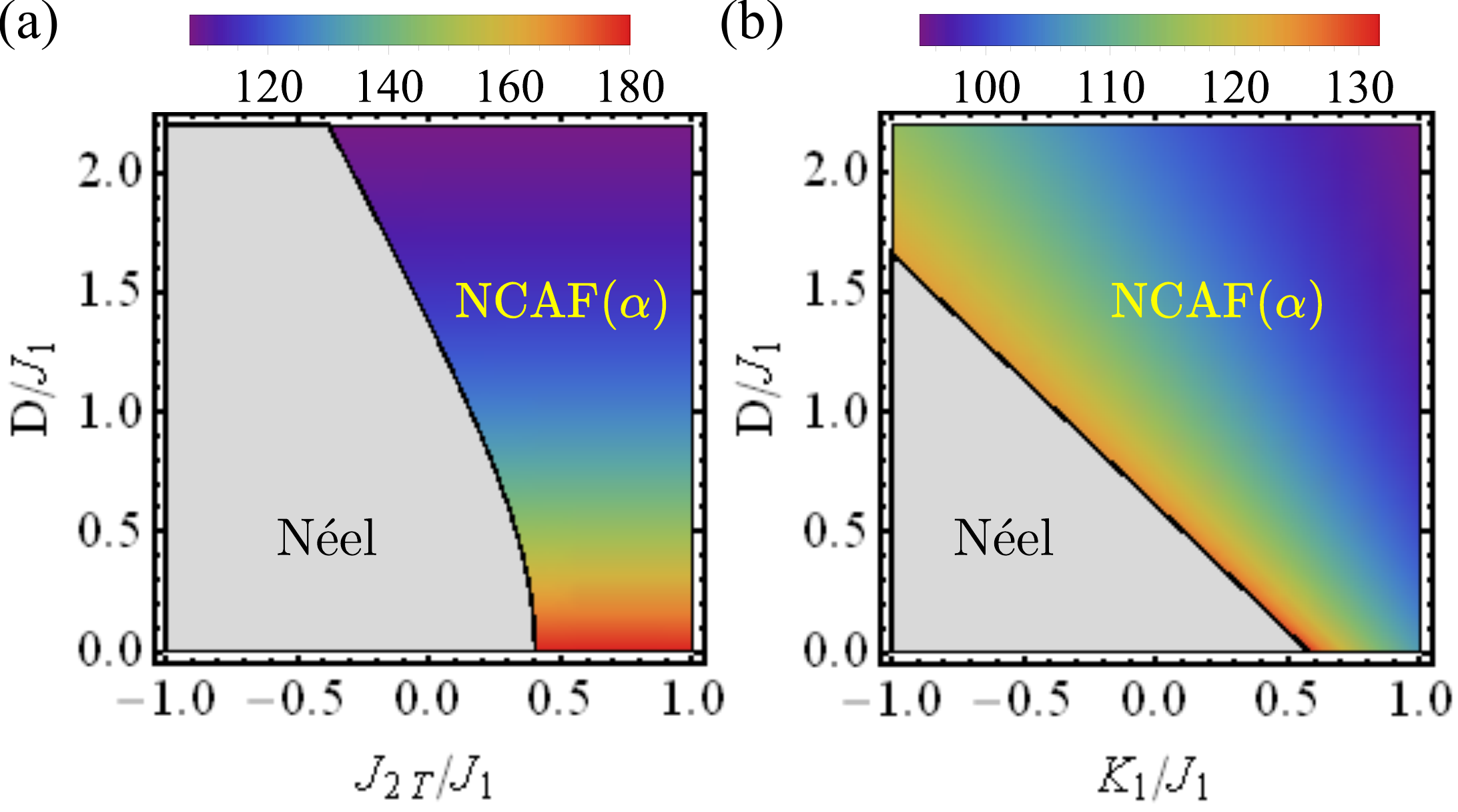}
    \caption{(a) The phase diagram as a function of parameters $D$ and ${J2}$ with the remaining exchange couplings fixed at the \textit{ab initio} values listed in Eq.~\eqref{eq:DFT} 
    and with fixed effective anisotropy parameters ($\gamma_{T}=30.41$, $\gamma_{O}=-0.53$). The false color denotes the optimized value of angle $\alpha$. (b) The phase diagram as a function of parameters $D$ and $K_1$, with the remaining exchange couplings fixed at the \textit{ab initio} values in Eq.~\eqref{eq:DFT}. 
    }
    \label{phasediagram3}
\end{figure}

So far, we have fixed the exchange parameters of the Hamiltonian to be those from the first principles calculations in Eq.~(\ref{eq:DFT}) and only varied the Dzyaloshinskii--Moriya interaction strength $D$.
Now, we relax the exchange parameters and investigate the phase diagram as a function of $J_{2T}/J_1$ and $D$ in Fig.~\ref{phasediagram3} (a). We see that the NCAF phase wins over the N\'eel phase provided $J_{2T}$ is sufficiently large, and the angle $\alpha$ varies continuously within the NCAF phase, shown as a false color in Fig.~\ref{phasediagram3}.
Similar conclusion is reached when we fix $J_{2T}$ to its \emph{ab initio} value and study the phase diagram as a function of the biquadratic interaction $K_1$, plotted in Fig.~\ref{phasediagram3} (b). In the latter case, the NCAF phase can be stabilized at an arbitrary value of $K_1$ (including $K_1=0$), provided $D$ is sufficiently large. Conversely, a large value of $K_1>0.39J_1$ favors the NCAF phase even in the absence of the DM interaction -- the same conclusion reached earlier in subsection~A (see Fig.~\ref{fig_phasediagram1}).

In Figures~\ref{phasediagram3} (a) and (b), the optimized angle $\alpha_{opt}$, shown as a false color, corresponds to the minimum NCAF energy under given $D$. This optimal angle decreases from $180^{\circ}$ (which corresponds to the collinear stripe phase, see Fig.~\ref{Fig_2d}c and \ref{Fig_2d}e) down to $90^{\circ}$ as $D$ increases, as expected since the larger DM interaction favors the noncollinear ordered state. 


\section{Quantum fluctuations around mean field}\label{sec:LSW}

In this section, we investigate the effect of quantum fluctuation around the mean-field solution for the ordered states. We perform linear spin wave theory (LSWT) calculations to compute the contribution of the magnon zero-point energy to the N\'eel and NCAF states, whose competition in Ni$_2$Mo$_3$O$_8$ is the principal goal of this work. Because LSWT can only handle the bilinear spin terms, at first we approximate the biquadratic-bilinear model to an effective Heisenberg model, which we then treat at the level of LSWT.

\subsection{Effective Heisenberg model} \label{subsec:effective_heisenberg}
Unless one is interested in quadrupolar spin ordering, which is not the case in Ni$_2$Mo$_3$O$_8$, it is often sufficient to approximate the biquadratic terms $(\vec{S}_i\cdot\vec{S}_j)^2$  by mean-field decoupling 
\begin{equation}
\label{MFdecoupling}
    J\vec{S}_i\cdot\vec{S}_j+K(\vec{S}_i\cdot\vec{S}_j)^2\approx J_{e}(\vec{S}_i\cdot\vec{S}_j)-K\langle\vec{S}_i\cdot\vec{S}_j\rangle^2
\end{equation}
whereby one obtains an effective Heisenberg model with an effective spin exchange $J_{e}=J+2K\langle\vec{S}_i\cdot\vec{S}_j\rangle$. However, it does not work well in the case of non-collinear ordering because the mean-field energy of the right hand side in Eq.~(\ref{MFdecoupling})   
\begin{equation}
    \langle J_{e}(\vec{S}_i\cdot\vec{S}_j)-K\langle\vec{S}_i\cdot\vec{S}_j\rangle^2\rangle=J\cos{\alpha_{ij}}+K\cos^2{\alpha_{ij}}
\end{equation}
is far from the expectation value of the energy of the left hand side computed quantum-mechanically for spin 1 objects:
\begin{equation}
    \langle J\vec{S}_i\cdot\vec{S}_j+K(\vec{S}_i\cdot\vec{S}_j)^2\rangle\!=\!
    \left(J-\frac{K}{2}\right)\cos{\alpha_{ij}}+\frac{K}{4}\cos^2{\alpha_{ij}}+\frac{5}{4}K
\end{equation}
except for when the two spins are alligned ferromagnetically ($\alpha_{ij}=0$). 

\begin{figure}[tbp]
    \centering
    \includegraphics[width=0.4\textwidth]{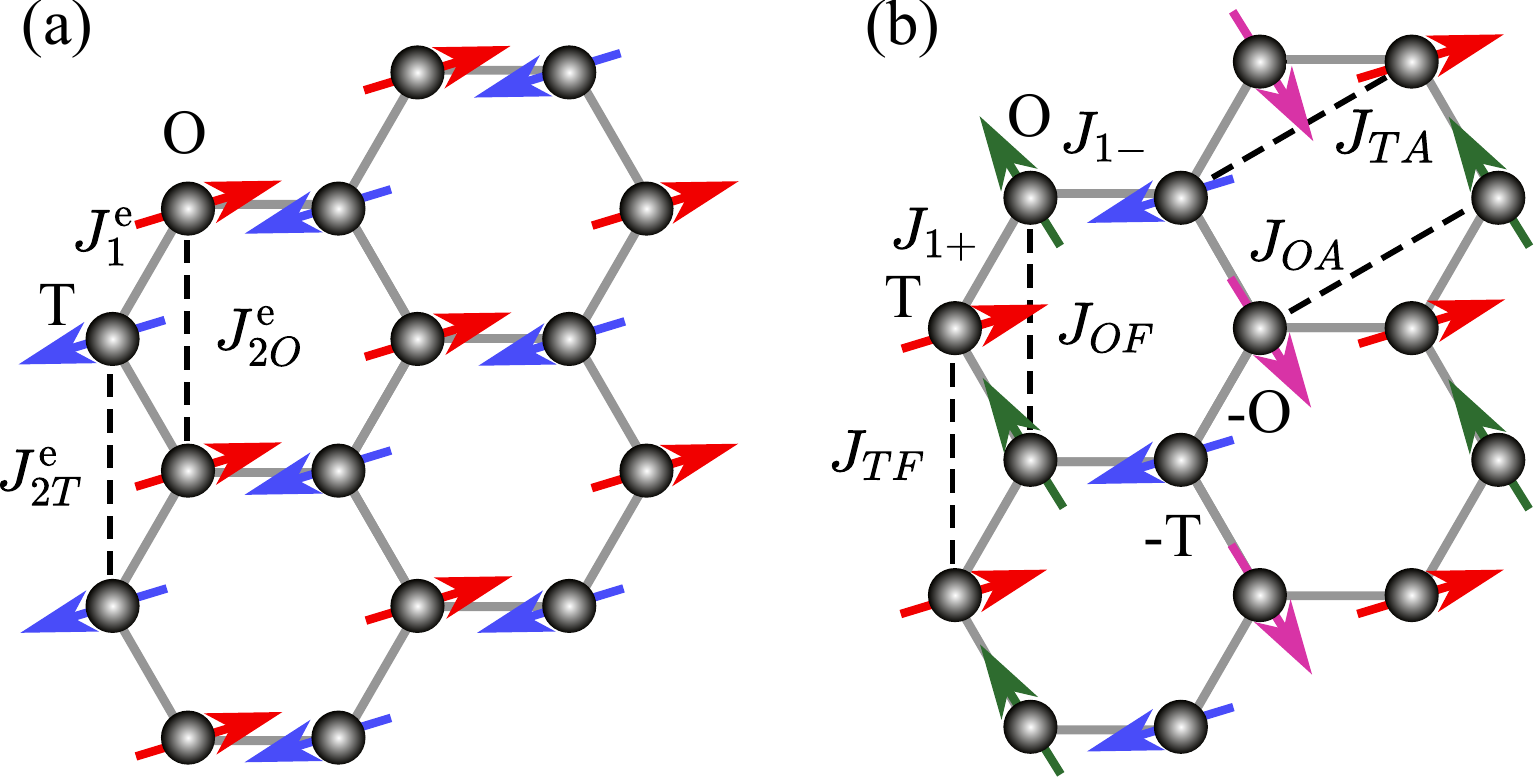}
    \caption{The depiction of the effective Heisenberg model for (a) N\'eel state and (b) NCAF state. The spins in the sketch of NCAF state are used for showing the effective model clearly, however, we should notice that the spins of NCAF state are in xz plane.}
     \label{fig:approxH}
\end{figure}

Instead, we approximate the spin-spin interaction to
\begin{equation}
    J\vec{S}_i\cdot\vec{S}_j+K(\vec{S}_i\cdot\vec{S}_j)^2\approx J_{e}(\vec{S}_i\cdot\vec{S}_j)-K\langle\vec{S}_i\cdot\vec{S}_j\rangle^2+f(K)
\end{equation}
with a constant $f(K)$ to be determined, by requiring that the expectation values of the energy on the two sides of the above equation are equal:
\begin{equation}
    J_e\langle\vec{S}_i\cdot\vec{S}_j\rangle-K\langle\vec{S}_i\cdot\vec{S}_j\rangle^2+f(K)=\langle J\vec{S}_i\cdot\vec{S}_j+K(\vec{S}_i\cdot\vec{S}_j)^2\rangle.
\end{equation}
This yields an \textit{effective} Heisenberg exchange coupling
\beq
J_e(\alpha_{ij})=J+\left(\frac{5}{4}\cos{\alpha_{ij}}-\frac{1}{2}\right)K
\eeq 
and $f(K)=5/4K$. Notice that the parameters of the resulting model explicitly depend on the angle $\alpha_{ij}$ between the spins in the ordered state. For example, 
for the N\'eel state shown in Fig.~\ref{fig:approxH}(a), the effective model reads
\begin{equation}
\label{eq:effneelmodel}
    \mathcal{H}_{\text{N\'eel}}^{\text{e-int}}=\sum_{\langle ij\rangle}J_{1}^{\text{e}}\vec{S}_i\cdot\vec{S}_j+\sum_{\langle\langle ij\rangle\rangle,T}J_{2T}^{\text{e}}\vec{S}_i\cdot\vec{S}_j+\sum_{\langle\langle ij\rangle\rangle,O}J_{2O}^{\text{e}}\vec{S}_i\cdot\vec{S}_j,
\end{equation}
with the effective coupling constants given by
\begin{equation}
\label{eq:effneelparameter}
    J_{1}^{\text{e}}=J_1-\frac{7}{4}K_1,~J_{2T}^{\text{e}}=J_{2T}+\frac{3}{4}K_{2T},~J_{2O}^{\text{e}}=J_{2O}+\frac{3}{4}K_{2O}.
\end{equation}

Similarly, the effective spin-bilinear Hamiltonian for the NCAF state shown in Fig.~\ref{fig:approxH}(b) takes the form
\begin{widetext}
\begin{eqnarray}
\label{eq:effncafmodel}
\mathcal{H}_{\text{NCAF}}^{\text{e-int}} & = &
\sum_{\langle T,O\rangle,\langle -T,-O\rangle}J_{1+} \vec{S}_{i}\cdot\vec{S}_{j} + \sum_{\langle T,-O\rangle,\langle O,-T\rangle}J_{1-} \vec{S}_{i}\cdot\vec{S}_{j} 
+\sum_{\langle\langle T,T\rangle\rangle}J_{TF} \vec{S}_{i}\cdot\vec{S}_{j}+\sum_{\langle\langle -T,-T\rangle\rangle}J_{TF} \vec{S}_{i}\cdot\vec{S}_{j}\\
& + &\sum_{\langle\langle T,-T\rangle\rangle}J_{TA} \vec{S}_{i}\cdot\vec{S}_{j}
+\sum_{\langle\langle O,O\rangle\rangle}J_{OF} \vec{S}_{i}\cdot\vec{S}_{j} +\sum_{\langle\langle -O,-O\rangle\rangle}J_{OF} \vec{S}_{i}\cdot\vec{S}_{j} 
+ \sum_{\langle\langle O,-O\rangle\rangle} J_{OA}\vec{S}_{i}\cdot\vec{S}_{j},
+\sum_{\langle ij\rangle}\vec{D}_{ij}\cdot(\vec{S}_i\times\vec{S}_j)\nonumber
\end{eqnarray}
\end{widetext}
where $-T$ corresponds to the tetrahedral sites with the spin direction $(\theta_{T}+\pi,\phi_{T})$ (denoted by blue arrows in Fig.~\ref{fig:approxH}b) and $-O$ denotes the octahedral site with the spin pointing along $(\theta_{O}+\pi,\phi_{O})$ (denoted by magenta arrows in Fig.~\ref{fig:approxH}b). Thus the price paid for writing down the effective Heisenberg model, is that the effective spin exchange couplings become anisotropic. The total number of effective coupling constants thus increases from 3 to 6, compared to the N\'eel state, with the values given by
\begin{equation}
\label{eq:effncafparameter}
\begin{aligned}
 &J_{1+}(\theta_T,\theta_O)=J_1+K_1\Big(\frac{5}{4}\cos(\theta_T+\theta_O)-\frac{1}{2}\Big),\\
 &J_{1-}(\theta_T,\theta_O)=J_1+K_1\Big(-\frac{5}{4}\cos(\theta_T+\theta_O)-\frac{1}{2}\Big),\\
 &J_{TF}=J_T+\frac{3}{4}K_T,\quad J_{TA}=J_T-\frac{7}{4}K_T,\\
 &J_{OF}=J_O+\frac{3}{4}K_O, \quad J_{OA}=J_O-\frac{7}{4}K_O.
\end{aligned}    
\end{equation}

\subsection{Linear spin wave theory}
Since the LSWT is the large-$S$ expansion around the ordered state, it is convenient to choose 
the local spin quantization axis $z_i$ on each site along the direction of the spin in the given state. In this local frame, the Holstein--Primakoff (HP) transformation of the spin operators  is given by the standard expressions
\begin{equation}
\label{eq:HPtrans}
    s^{+}_{i}=\sqrt{2S}b,~s^{-}_{i}=\sqrt{2S}b^{\dagger},~s^{z}_{i}=S-b^{\dagger}b.
\end{equation}
The original spin operators $S_i$ in the laboratory frame are related to these local spin operators $s_i$ by a rotation in  
the $xz$ plane as follows:
\begin{equation}
    \begin{aligned}
        S_i^x&=\cos{\theta_i}\,s^{x}_{i}-\sin{\theta_i}\,s^{z}_{i},\\
        S_i^y&=s^{y}_{i},\\
        S_i^z&=\sin{\theta_i}\,s^{x}_{i0}+\cos{\theta_i}\,s^{z}_{i}.
    \end{aligned}
\end{equation}
The Heisenberg interaction, expressed in terms of the HP bosons, thus becomes 
\begin{equation}\label{lswtrans}
    \begin{aligned}
       \vec{S}_i\cdot\vec{S}_j&=S^2\cos{\alpha_{ij}}-S\cos{\alpha_{ij}}(b_i^{\dagger}b_i+b_j^{\dagger}b_j)\\
       &+\frac{S}{2}(1+\cos{\alpha_{ij}})(b_i^{\dagger}b_j+b_ib_j+\text{h.c.})\\
       &+\sqrt{\frac{S}{2}}\sin{\alpha_{ij}}(b_i-b_j+\text{h.c.}),
    \end{aligned}
\end{equation}
and similarly for the DM interaction. After HP transformation, the effective Hamiltonian in momentum space can be written as
\begin{equation}\label{eq:HSW}
    \mathcal{H}=\mathcal{E}_0+\mathcal{E}_{C}+H_\text{lin}[b^{\dagger},~b]+\sum_{\vk} \psi_\vk^{\dagger}H(\vec{k})\psi_\vk,
\end{equation}
where $\mathcal{E}_0$ is the MF energy, $\mathcal{E}_{C}$ is the term from commutation relation when we construct bosonic Nambu representation written in terms of $\psi_\vk = [b_{T,\vk},b_{O,\vk},b^\dagger_{T,-\vk},b^\dagger_{O,-\vk}]^{\intercal}$, where $b_T$ and $b_O$ are annihilation operators at T and O sites, respectively. The explicit form of the matrix $H(\vk)$ is shown in Appendix~\ref{app:Spinwave}. Above, $H_\text{lin}$ is the part of the Hamiltonian linear in the boson creation/annihilation operators, which we ignore as it does not conserve the number of bosons (magnons). Physically, this term appears when the reference magnetic state is not the saddle-point of the Hamiltonian, which may happen in the NCAF state for technical reasons to do with approximating the biquadratic spin interaction via an effective Heisenberg term. 

The last term in Eq.~\eqref{eq:HSW},
after the Bogoliubov transformation, becomes diagonal in the Bogoliubov operator basis, resulting in the zero-point fluctuation contribution to the energy of an ordered state:
\begin{equation}
    \mathcal{E}_{LSW}=\mathcal{E}_0+\mathcal{E}_{C}+\frac{3\sqrt{3}}{4(2\pi)^2}\sum_i\sum_{\vk \in BZ}E_i(\vec{k}),
\end{equation}
where $E_i(\vec{k})$ labels the positive eigenenergies of $H(\vk)$ and the sum is over all the magnon bands. We perform the linear spin wave calculation for the two competing states: N\'eel and NCAF, see Appendix~\ref{app:Spinwave} for details.

\subsection{LSWT Result}
Since there is no coupling between the layers in our model (it is believed to be very small in Ni$_2$Mo$_3$O$_8$~\cite{nafexp}), our LSWT calculation are effectively two-dimensional. We note that the energy of the NCAF state depends on both angles $\alpha$ and $\theta_{T}$. Under the fixed anisotropy parameters (see section~\ref{sec:SIA}) and exchange couplings determined from \textit{ab initio} calculations (section~\ref{sec:dft}), we vary the strength of the Dzyaloshinskii--Moriya interaction $D$  and optimize the angles $\alpha$ and $\theta_{T}$ to obtain the minimum energy of the NCAF state.

\begin{figure}[htbp]
    \centering
    \includegraphics[width=0.45\textwidth]{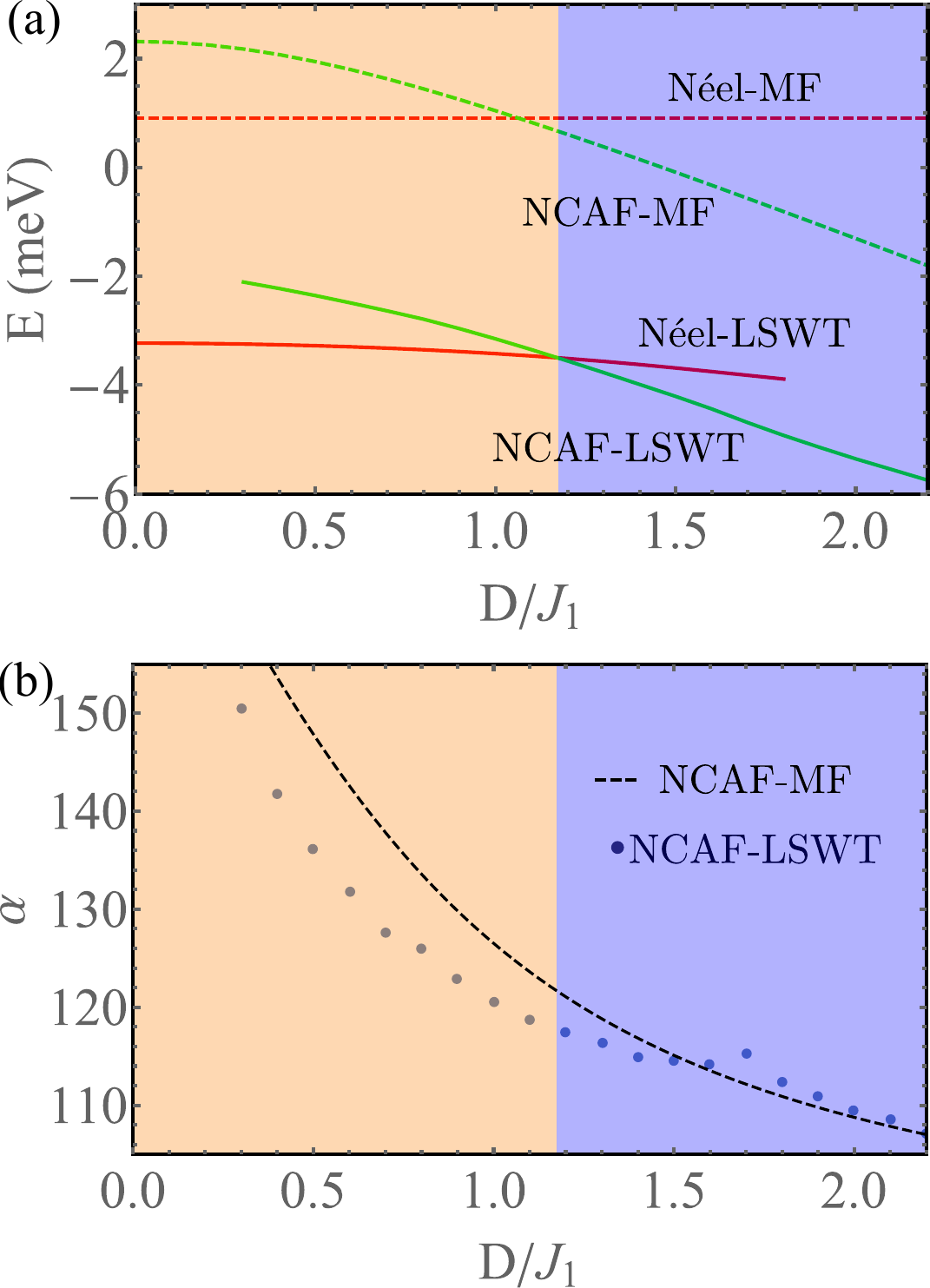}
    \caption{(a) Energy of N\'eel and NCAF state as functions of DM strength $D$. Solid red (green) line corresponds to N\'eel (NCAF) states from LSWT. Dashed red (green) line corresponds to the MF energy of N\'eel (NCAF) states. (b) Angle $\alpha$ between T and O sites at different $D$. The dotted (dahsed) line is from LSWT (MF). The orange region corresponds to N\'eel phase, where $\alpha=180^\circ$, whereas $\alpha$ is $D$-dependent in the blue NCAF phase.}
    \label{phasediagram5}
\end{figure}

The resulting phase diagram is shown in Fig.~\ref{phasediagram5}(a). The energy of the N\'eel (NCAF) state is represented by solid red (green) line, respectively. For comparison, the MF energies of these two state are represented by the dashed lines of the same colors. After we consider the zero-point fluctuation, the energies of both the N\'eel and NCAF states decrease compared with MF result. 
The phase boundary between the two phases $D_c^{LSW}\approx1.175J_1$ changes slightly from the MF result $D_c^{MF}\approx1.06J_1$, which does not qualitatively affect any of our conclusions. In Fig.~\ref{phasediagram5}(b), we plot the optimal angle $\alpha$ between the spins on the T and O sites inside the NCAF phase as a function of DM interaction $D$, with the dotted (dashed) line corresponding  to the LSWT (MF) results, respectively. Of course the angle $\alpha=180^\circ$ for $D<D_c$ inside the N\'eel phase, so the plotted value of $\alpha$ is only meaningful on the right-hand side of the boundary where the NCAF phase becomes stable.
Right at the phase boundary the angle $\alpha\approx115^{\circ}$, and its value decreases almost monotonically with increasing $D$, except for an anomaly near $D/J_1=1.7$.  

\begin{figure}[t]
    \centering
    \includegraphics[width=0.9\columnwidth,viewport=92 238 520 554]{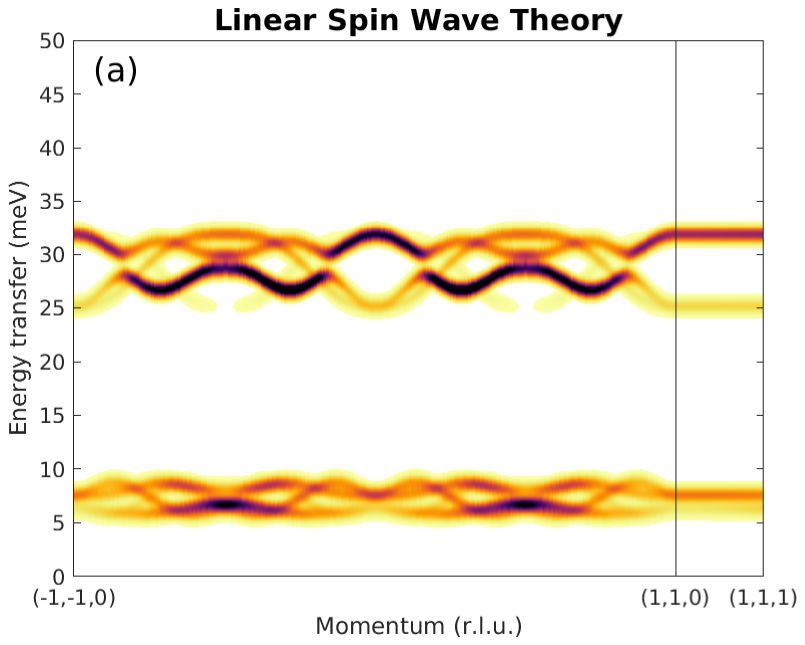}
    \includegraphics[width=0.9\columnwidth,viewport=92 238 520 554]{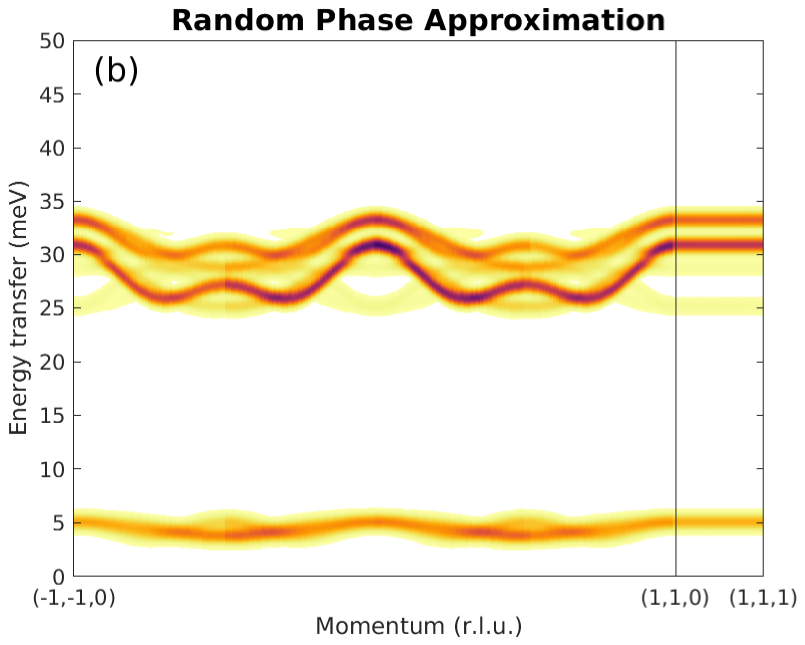}
    \caption{Calculated magnetic excitation spectra using either (a) linear spin wave theory treating only the spin degree of freedom, or (b) treating the full crystal field effects in a random phase approximation calculation.}
    \label{fig:spectra}
\end{figure}

In Figure~\ref{fig:spectra}, we present the calculated excitation spectra using the spin wave theory outlined above in a full lattice (with two T and two O sites per unit cell which contains two honeycomb layers). The upper panel (a) shows the calculated magnetic excitation spectrum with a spin-only linear spin wave theory using the single-ion anisotropy parameters $\gamma_T$ and $\gamma_O$ given in Section~\ref{sec:SIA} and the DFT-derived exchange couplings from Eq.~\eqref{eq:DFT}. The lower panel (b) shows the a calculation under the random phase approximation (RPA) which includes the full crystal field Hamiltonian. In both cases the  DM interaction strength was fixed at $D=3.1$~meV. The \emph{SpinW}~\cite{spinw} program was used for the calculations in Fig.~\ref{fig:spectra}(a) whilst \emph{McPhase}~\cite{mcphase} was used for those in panel (b). 

Both calculations give two bands of excitations around 5 and 25~meV. The lower energy band is from the octahedral sites, whilst the higher energy excitations is associated with the tetrahedral sites, which have a larger single-ion anisotropy parameter $\gamma_T$ in Eq.~(\ref{eq:anisotropy}). The upper band dispersion is similar in the two calculations albeit the modes have different relative intensities, but the lower bands differ qualitatively. The differences arises due to the way the two calculation methods treat the single-ion anisotropy, with the RPA theory being more accurate, as described in Appendix~\ref{app:Spinwave_spectrum}.


\section{Discussion} \label{sec:discussion}
In this work, we have proposed an effective spin model including the nearest neighbor Dzyaloshinskii--Moriya interaction to explain the noncollinear magnetic ordered state observed in a non-centrosymmetric honeycomb lattice material Ni$_2$Mo$_3$O$_8$~\cite{nafexp}. The reason for introducing the DM interaction is that it favors two neighboring spins to be perpendicular to each other, and competes with the bilinear $\vec{S}_i\cdot\vec{S}_j$ and biquadratic $(\vec{S}_i\cdot\vec{S}_j)^2$ terms, which usually favor two neighboring spins to be collinear (unless the biquadratic term is positive and large, see the discussion around Eq.~(\ref{eq:alpha-opt}), which is however not realized in Ni$_2$Mo$_3$O$_8$). We show that without the DM interaction, the purely bilinear, or bilinear-biquadratic model cannot reproduce the non-collinear magnetic order observed in Ni$_2$Mo$_3$O$_8$~\cite{nafexp}.

We further argue that considering the \textit{nearest neighbor} Dzyaloshinskii--Moriya interaction is sufficient. From Moriya's rules, the second-neighbor DM vector between two tetrahedral (T) or two octahedral (O) Ni spins lives in the plane which bisect the T-T (O-O) bond and is perpendicular to it. However, in both collinear states and NCAF state, the spins at next-nearest neighbor are collinear. As a result, the next-nearest neighbor DM interaction does not affect their MF energies, and gives only a small correction to the linear spin-wave theory. 
It is furthermore difficult to imagine the spins in the same sublattice (T or O) to be noncollinear, given that the crystal field environment and the magnetic anisotropy are the same on the two sites, corroborating the above conclusion that the n.n.n. DM interaction, even if present, does not contribute to the energies of the two competing states (N\'eel and NCAF).
As for the third-neighbor and longer-range DM interactions, those are expected to be negligible, given the large separation between the magnetic moments. 

In this work, we also considered the effect of single-ion anisotropy, following the detailed crystal field analysis (see Section~\ref{sec:SIA}). Although the crystal field environments on T and O sites are different, it does not lead to noncollinear spin order. Moreover, because the crystal-field parameter $\gamma_T\gg\gamma_O$, the T spins prefer to lie close to the $xy$ plane and in the absence of the DM interaction, remain collinear with the O spins.


In the experimental paper~\cite{nafexp}, several tentative scenarios were advanced to explain the noncollinear magnetic ordering in Ni$_2$Mo$_3$O$_8$. One of them was bond-dependent Kitaev-like interaction, however for it to be realized, the usual pathway is in systems with edge shared octahedral environment~\cite{HigherSpinKitaev}, which is not the case in Ni$_2$Mo$_3$O$_8$. Another possibility is that of a spiral state, which typically requires the exchange couplings $J_1, J_2, J_3$ up to third nearest neighbors to all have similar magnitude. This is however not the conclusion we have reached from our \textit{ab initio} calculations, where we find $J_3$ ($\sim10^{-2}$ meV) to be negligible. Finally, it was proposed~\cite{nafexp} that bond-dependent anisotropic interactions, through ligand distortion, may be the cause of the noncollinear magnetic order to appear in Ni$_2$Mo$_3$O$_8$. While we cannot exclude this latter mechanism, we would argue that the Dzyaloshinskii--Moriya interaction provides a more natural explanation and, as our results demonstrate (see Fig.~\ref{fig_phasediagram1} and Fig.~\ref{phasediagram3}), the optimal value of the angle $\alpha$ between neighboring spins is predicted to be close to the experimental value $\alpha'=130^\circ$~\cite{nafexp}.


In summary, we have demonstrated that the NCAF ordered states in Ni$_2$Mo$_3$O$_8$ can be successfully understood as stemming from the first neighbor Dzyaloshinskii--Moriya interaction. Using a combination of first principles electronic structure and product states' expeected energy calculations, we have estimated the values of the exchange couplings, established the expected energy phase diagram and found that a realistic value of DM interaction $D>D_c\approx 2.78$~meV is sufficient to stabilize the noncollinear magnetic order with the angle $\alpha_{\text{opt}}$ between the neighboring spins within a few degrees of the experimental value $\alpha'=130^\circ$. 

We have performed the linear spin-wave calculations to include the fluctuations around the saddle-point solutions, and found that the inclusion of zero-point energies does not qualitatively affects the main conclusion, only shifting the critical value of DM interaction imperceptibly. We further make predictions for the magnon spectra inside the noncollinear magnetic phase, which should be compared to future inelastic neutron scattering data on Ni$_2$Mo$_3$O$_8$.
Our calculations also indicate that when choosing between the two neutron scattering refinement fits reported in Ref.~\onlinecite{nafexp} and summarized in Table~\ref{table1}, the first fit with the angle $\alpha'=130^\circ$ receives support from both the \textit{ab initio} results and our theoretical calculations. 

The present study opens up a new exciting avenue for investigating frustrated spin-1 systems with spin-orbit induced Dzyaloshinskii--Moriya interactions. 
Application of the present ideas to different materials and lattices other than the honeycomb certainly deserve further attention.



\section{Acknowledgements}
The authors thank Tyrel McQueen for many fruitful discussions. This work was supported by the Robert A. Welch Foundation Grant No.~C-1818. A.H.N. also acknowledges the support of the National Science Foundation Division of Materials Research under the Award DMR-1917511. S.L. and A.H.N. acknowledge the hospitality of the Kavli Institute for Theoretical Physics (supported by the NSF Grant No. PHY-1748958), where a portion of this work was performed.


\appendix
\section{Mean field energy of various states}\label{app:MF}

The total Hamiltonian for spin $S=1$ is
\begin{equation}
\begin{aligned}
\mathcal{H}_\text{eff} &= 
\sum_{\langle ij\rangle}J_{1} \vec{S}_{i}\cdot\vec{S}_{j} + K_{1} (\vec{S}_{i}\cdot\vec{S}_{j})^2 \\
&+\sum_{\langle\langle ij\rangle\rangle,T}J_{2T} \vec{S}_{i}\cdot\vec{S}_{j} + K_{2T} (\vec{S}_{i}\cdot\vec{S}_{j})^2\\
&+\sum_{\langle\langle ij\rangle\rangle,O}J_{2O} \vec{S}_{i}\cdot\vec{S}_{j} + K_{2O} (\vec{S}_{i}\cdot\vec{S}_{j})^2\\
&+\sum_{\langle ij\rangle}\vec{D}_{ij}\cdot(\vec{S}_i\times\vec{S}_j)\\
&+\sum_{i,T}\gamma_T(\vec{S}_i^z)^2+\sum_{i,O}\gamma_O(\vec{S}_i^z)^2,\\
\end{aligned}
\end{equation} 
where the indices T and O denote the tetrahedral and octahedral Ni sites, respectively.

Based on MF ansatz in Eq.~(\ref{eq:productstate}) and Eq.~(\ref{eq:spinstate}), we consider the interactions between two spins $\vec{S}_i$ and $\vec{S}_j$ parametrized by polar and azimuthal angles $\theta_i$, $\phi_i$ and $\theta_j$, $\phi_j$, respectively. The MF expressions of the terms in the Hamiltonian take the following form:
\begin{equation}\label{MFTeq}
\begin{aligned}
\langle\vec{S}_{i}\cdot\vec{S}_{j}\rangle=&\cos\alpha_{ij},\\
\langle(\vec{S}_{i}\cdot\vec{S}_{j})^2\rangle=&\frac{1}{4}\cos^2\alpha_{ij}-\frac{1}{2}\cos\alpha_{ij}+\frac{1}{4}+1,\\
\langle\vec{S}_{i}\times\vec{S}_{j}\rangle=&\sin\theta_{i}\sin\phi_{i}\cos\theta_{j}-\sin\theta_{j}\sin\phi_{j}\cos\theta_{i}\\
+&\sin\theta_{j}\cos\phi_{j}\cos\theta_{i}-\sin\theta_{i}\cos\phi_{i}\cos\theta_{j}\\
+&\sin\theta_{i}\cos\phi_{i}\sin\theta_{j}\sin\phi_{j}\\
-&\sin\theta_{i}\sin\phi_{i}\sin\theta_{j}\cos\phi_{j},\\
\langle(\vec{S}_i^z)^2\rangle=&\frac{1}{2}\cos^2\theta_i+\frac{1}{2},
\end{aligned}
\end{equation}
where $\alpha_{ij}$ is the angle between two spin directions, $\cos\alpha_{ij}=\sin\theta_{i}\sin\theta_{j}\cos(\phi_{i}-\phi_{j})+\cos\theta_{i}\cos\theta_{j}$. 
For simplicity, we get rid of a constant 1 in $\langle(\vec{S}_{i}\cdot\vec{S}_{j})^2\rangle$ term. With these mean field results, we can obtain the average energy per site of the FM, N\'eel, stripe, zigzag and NCAF state quoted in Eqs.~(\ref{ewodm}) and (\ref{eq:E-DM-NCAF}):

\begin{equation}
\label{eq:mftenergyinappdix}
\begin{aligned}
\mathcal{E}_{\text{FM}}^*=&\frac{3}{2}J_1+\frac{3}{2}J_{2T}+\frac{3}{2}J_{2O},\\
\mathcal{E}_{\text{N\'eel}}^*=&-\frac{3}{2}J_1+\frac{3}{2}K_1+\frac{3}{2}J_{2T}+\frac{3}{2}J_{2O},\\
\mathcal{E}_{\text{Stripe}}^*=&-\frac{1}{2}J_1+K_1-\frac{1}{2}J_{2T}+K_{2T}-\frac{1}{2}J_{2T}+K_{2O},\\
\mathcal{E}_{\text{Zig-Zag}}^*=&\frac{1}{2}J_1+\frac{1}{2}K_1-\frac{1}{2}J_{2T}+K_{2T}-\frac{1}{2}J_{2T}+K_{2O},\\
\mathcal{E}_{\text{NCAF}}^*=&\frac{1}{2}J_1\cos\alpha+K_1(\frac{3}{8}\cos^2\alpha-\frac{1}{4}\cos\alpha+\frac{3}{8})\\
&-\frac{1}{2}J_{2T}+K_{2T}-\frac{1}{2}J_{2O}+K_{2O}\\
&-D(\sin\theta_O\cos\phi_O\cos\theta_T-\sin\theta_T\cos\phi_T\cos\theta_O),\\
\end{aligned}
\end{equation}
where we use the asterisk (*) to label the energy without single ion anisotropy. The contribution to the energy from the anisotropy term 
\begin{equation}
    \mathcal{E}_A(\theta_T,\theta_O)=\frac{1}{4}\gamma_T(\cos^2\theta_T+1)+\frac{1}{4}\gamma_O(\cos^2\theta_O+1)
\end{equation}

We notice that the single ion anisotropy is poorly captured by our \emph{ab initio} DFT calculation, thus we can only solve six exchange parameters $J_1$, $J_{2T}$, $J_{2O}$, $K_1$, $K_{2T}$ and $K_{2O}$. The first four of the above equations are linearly dependent, and we therefore need at least three other noncollinear states in order to be able to solve for these six parameters. To make the result more accurate, we have increased the number of the reference states to 12 and perform least squares fitting to obtain the exchange parameters. The other 8 states are as follows.
 
 \begin{figure}[tbh!]
    \centering
    \includegraphics[width=0.42\textwidth]{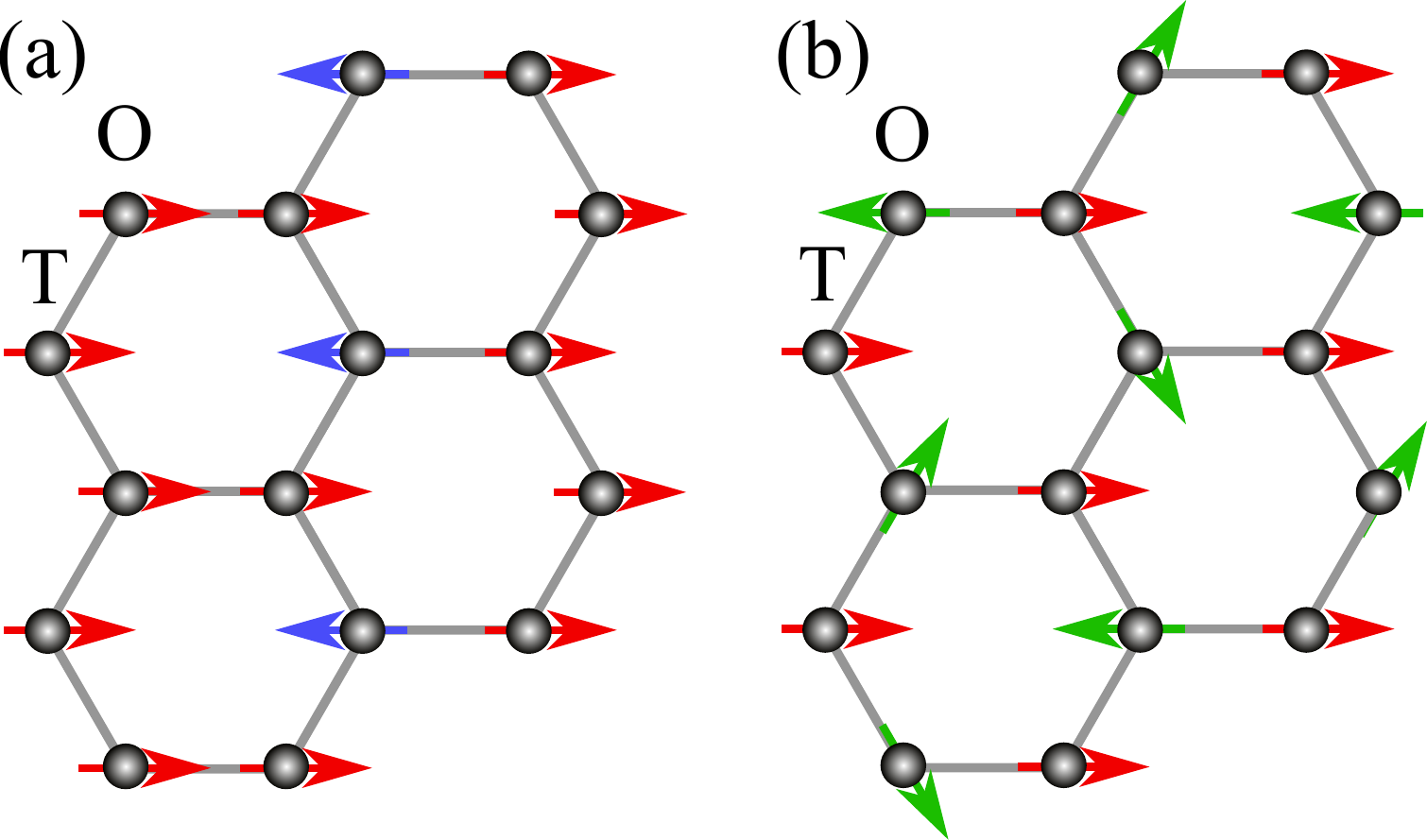}
    \caption{The depiction of (a) a collinear state with ferromagnetic order in T sites and antiferromagnetic order in O sites and (b) a noncollinear state with ferromagnetic order in T sites and $120^{\circ}$ antiferromagnetic order in O sites.}
    \label{fig:newstate}
\end{figure}
 
First we consider two collinear states which have ferromagnetic order in T (O) sites and antiferromagnetic order in O (T) sites, as shown in Fig.~\ref{fig:newstate} (a). The energy expressions are
\begin{equation}
\begin{aligned}
\mathcal{E}_1^*=&\frac{3}{4}K_1+\frac{3}{2}J_{2T}-\frac{1}{2}J_{2O}+K_{2O},\\
\mathcal{E}_2^*=&\frac{3}{4}K_1+\frac{3}{2}J_{2O}-\frac{1}{2}J_{2T}+K_{2T}.\\
\end{aligned}
\end{equation}
Then we rotate the spins at sublattice with ferromagnetic order by $90^{\circ}$, there are two new collinear states:
\begin{equation}
\begin{aligned}
\mathcal{E}_3^*=&\frac{3}{8}K_1+\frac{3}{2}J_{2T}-\frac{1}{2}J_{2O}+K_{2O},\\
\mathcal{E}_4^*=&\frac{3}{8}K_1+\frac{3}{2}J_{2O}-\frac{1}{2}J_{2T}+K_{2T}.\\
\end{aligned}
\end{equation}  
Besides that, we introduce two noncollinear analogues of the N\'eel and zigzag states, obtained by rotating the spins on one of the sublattices (say, blue) in Figs.~\ref{Fig_2d}d) and \ref{Fig_2d}f) respectively, such that the spins on the red and blue sublattice are perpendicular to each other. The mean-field energies of these two states are
\begin{equation}
\label{ncenergy}
\begin{aligned}
\mathcal{E}_{5}^*=&\frac{3}{8}K_1+\frac{3}{2}J_{2T}+\frac{3}{2}J_{2O}\\
\mathcal{E}_{6}^*=&J_1+\frac{1}{8}K_1+\frac{1}{2}J_{2T}+\frac{1}{4}K_{2T}+\frac{1}{2}J_{2O}+\frac{1}{4}K_{2O}.
\end{aligned}
\end{equation}
Finally, we consider two noncollinear states with ferromagnetic order in T (O) sites and $120^{\circ}$ antiferromagnetic order in O (T) sites, as shown in Fig.~\ref{fig:newstate} (b). The mean-field energies of these two states are
\begin{equation}
\begin{aligned}
\mathcal{E}_7^*=&\frac{3}{8}K_1+\frac{3}{2}J_{2T}-\frac{3}{4}J_{2O}+\frac{27}{32}K_{2O},\\
\mathcal{E}_8^*=&\frac{3}{8}K_1+\frac{3}{2}J_{2O}-\frac{3}{4}J_{2T}+\frac{27}{32}K_{2T}.\\
\end{aligned}
\end{equation}

For all these 12 states, we avoid the Dzyaloshnskii--Moriya interaction, since DFT has difficulty accurately capturing those. With mean field and DFT results of these reference states, we perform least-square fitting to  minimize the discrepancies between the analytical and \textit{ab initio} energy differences of the references states: 
\begin{equation}
    \sum_i\Big((\mathcal{E}_i^*(J,K)-\mathcal{E}_{FM}^*(J,K))-(\mathcal{E}_i^{DFT}-\mathcal{E}_{FM}^{DFT})\Big)^2
\end{equation}
under $|J_i|>|K_i|$, $J_{TF}-2J_{TA}<0$ and $J_{OF}-2J_{OA}<0$, the last two are weak constrains that stabilize the collinear antiferromagtic ordered spins in two sublattices, $J_{TF},~J_{TA},~J_{OF},~J_{OA}$ are effective Heisenberg exchange couplings introduced in section~\ref{sec:LSW}. This yields the values of the exchange parameters $J_1,K_1,J_{2T},K_{2T},J_{2O}$ and $K_{2O}$ listed in Eq.~\eqref{eq:DFT} in the main text with a very good fit-quality factor $R^2=0.956$. The comparison between the \textit{ab initio} and the resulting model energies is shown in Fig.~\ref{fig:fit} in the main text.



\section{Details of \emph{ab initio} analysis}\label{app:DFT}

\begin{figure}[t]
    \centering
    \includegraphics[width=0.5\textwidth]{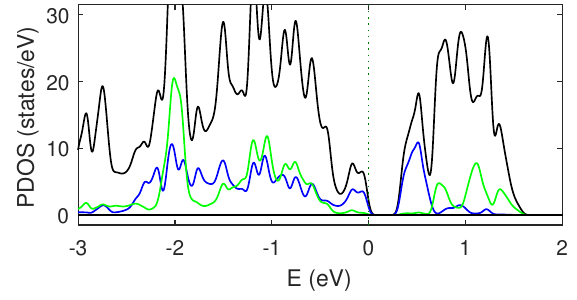}
    \caption{The total electronic density of state (black line) in the N\'eel state of Ni$_2$Mo$_3$O$_8$. The partial contributions from 
    Ni$_T$ (green) and 
    Ni$_O$ (blue) electrons are also shown. The calculation was performed within DFT, with the Hubbard parameter $U=0$. A clear band gap appears at the chemical potential, indicating that the magnetically ordered compound is a band insulator.}
    \label{fig_dos}
\end{figure}

We performed the first-principles DFT+U~\cite{dftu} calculations as implemented in the VASP package\cite{vasp} using the projector augmented wave method\cite{paw}, making use of GGA-PBE for exchange-correlation potential\cite{pbe}. In order to extract the parameters of the effective spin model, we considered various possible magnetically ordered states, including ferromagnet (FM), N\'eel, stripe, zigzag as well as the noncollinear NCAF state.
The base unit cell consisting of 2 stacked layers, with 2 Ni atoms per layer is sufficient to describe the the FM and N\'eel states. A $2\times1\times1$ supercell is used to describe the stripe and zigzag states, while a $2\times2\times1$ supercell is used to describe the NCAF state. We have performed the calculations without and with the Hubbard interaction $U=3$ eV. The moments reported in section~\ref{sec:dft} in the main text were calculated for $U=0$. On increasing the value of the Hubbard $U$ to $U=3$~eV, the total moments increased by about $0.2\mu_B$. The rest of our conclusions remain unchanged.

The density of states plots reveal the insulating nature of the compound, as shown in Fig.~\ref{fig_dos}. There is an insulating gap even at Hubbard $U=0$, which widens further with $U=3$ eV. This suggests that Ni$_2$Mo$_3$O$_8$ is a Slater insulator, with the gap opening due to magnetism, rather than due to the Hubbard on-site repulsion.

\section{Classical ground state without spin orbit coupling}\label{app:LT}

In the main text, we consider the competition between two states -- the N\'eel phase and the noncollinear antiferromagnet (NCAF), see Fig.~\ref{fig_phasediagram1}. This is justified \textit{a posteriori} by comparing the energies of the two phases both at the mean-field level (section \ref{sec:MFT}) and with zero-point fluctuations taken into account (section~\ref{sec:LSW}). Here, we provide an alternative, unbiased proof that the N\'eel antiferromagnet is indeed the classical ground state of the model in Eq.~(\ref{HwithoutDM}), before the DM interactions are taken into account.

We apply the classical Luttinger–Tisza method~\cite{LT1,LT2} to find the ground state of our model under the fitting parameters in Eq.~(\ref{eq:DFT}), without DM interaction. We also ignore the single ion anisotropy, since it lies beyond the classical approach described below.

First we notice that the biquadratic spin interaction can be transformed into quadrupolar interaction
\begin{equation}
    (\vec{S}_i\cdot\vec{S}_j)^2=\frac{1}{2}\vec{Q}_i\cdot\vec{Q}_j-\frac{1}{2}\vec{S}_i\cdot\vec{S}_j+\frac{1}{3}S^2(S+1)^2,
\end{equation}
where 5 linearly independent quadrupolar components are
\begin{equation}
  \begin{pmatrix}
  Q^{x^2-y^2}\\
  Q^{3z^2-r^2}\\
  Q^{xy}\\
  Q^{yz}\\
  Q^{zx}
  \end{pmatrix}
  =
  \begin{pmatrix}
  (S^x)^2-(S^y)^2\\
  \frac{1}{\sqrt{3}}[3(S^z)^2-S(S+1)]\\
  S^xS^y-S^yS^x\\
  S^yS^z-S^zS^y\\
  S^zS^x-S^xS^z
  \end{pmatrix}.
\end{equation}
The quadrupolar and dipolar components are not independent, they satisfy a strong constraint  $(\vec{S}_i)^2+(\vec{Q}_i)^2=\frac{4}{3}$ on each site. In order to account for the possibility of both the dipolar and quadrupolar order, we introduce a constant $0\leq\eta\leq 1$ ($\eta=1$ would correspond to a purely dipolar state), in terms of which we have two constraints:
\begin{equation}
\begin{aligned}
    \frac{1}{\beta_{1}^2}\sum_{i,T}(\vec{S}_i)^2+\frac{1}{\beta_{2}^2}\sum_{i,O}(\vec{S}_i)^2&=\eta \frac{N}{2}(\frac{1}{\beta_{1}^2}+\frac{1}{\beta_{2}^2})\\
    \frac{1}{\beta_{3}^2}\sum_{i,T}(\vec{Q}_i)^2+\frac{1}{\beta_{4}^2}\sum_{i,O}(\vec{Q}_i)^2&=(\frac{4}{3}-\eta) \frac{N}{2}(\frac{1}{\beta_{3}^2}+\frac{1}{\beta_{4}^2})\\
    \label{eq:constraints}
\end{aligned}
\end{equation}
where $\beta_{1},\beta_3$ (and $\beta_2, \beta_4$) are any real, non-zero numbers that capture the relative contribution of dipolar and quadrupolar moments on the T site (O site), respectively. Here $N$ is the total number of sites. 
Introducing the Lagrange mutlipliers to enforce the above constraints on average, we obtain the following Lagrangian function, to be minimized
\begin{equation}
\begin{aligned}
\mathcal{L}&= 
\sum_{\langle ij\rangle}(J_{1}-\frac{K_{1}}{2}) \vec{S}_{i}\cdot\vec{S}_{j} + \frac{K_{1}}{2} (\vec{Q}_{i}\cdot\vec{Q}_{j})^2+\frac{4}{3}K_{1}\\
+&\sum_{\langle\langle ij\rangle\rangle,T}(J_{2T}-\frac{K_{2T}}{2}) \vec{S}_{i}\cdot\vec{S}_{j} + \frac{K_{2T}}{2} (\vec{Q}_{i}\cdot\vec{Q}_{j})^2+\frac{4}{3}K_{2T}\\
+&\sum_{\langle\langle ij\rangle\rangle,O}(J_{2O}-\frac{K_{2O}}{2}) \vec{S}_{i}\cdot\vec{S}_{j} + \frac{K_{2O}}{2} (\vec{Q}_{i}\cdot\vec{Q}_{j})^2+\frac{4}{3}K_{2O}\\
-&\lambda^S\left(\frac{1}{\beta_{1}^2}\sum_{i,T}(\vec{S}_i)^2+\frac{1}{\beta_{2}^2}\sum_{i,O}(\vec{S}_i)^2-\eta \frac{N}{2}(\frac{1}{\beta_{1}^2}+\frac{1}{\beta_{2}^2})\right)\\
-&\lambda^Q\left(\frac{1}{\beta_{3}^2}\sum_{i,T}(\vec{Q}_i)^2+\frac{1}{\beta_{4}^2}\sum_{i,O}(\vec{Q}_i)^2-(\frac{4}{3}-\eta) \frac{N}{2}(\frac{1}{\beta_{3}^2}+\frac{1}{\beta_{4}^2})\right).\\
\end{aligned}
\end{equation} 
The minimum (more generally, saddle point) of this function  satisfies the equations:
\begin{equation}
    \frac{\partial\mathcal{L}}{\partial\lambda^S}=0, ~\frac{\partial\mathcal{L}}{\partial\lambda^Q}=0,
    ~\frac{\partial\mathcal{L}}{\partial S_i^{\alpha}}=0,
    ~\frac{\partial\mathcal{L}}{\partial Q_i^{\gamma}}=0,
\end{equation}
where the index $\alpha=x,y,z$ labels the spin components on a given site and $\gamma=x^2-y^2,3z^2-r^2,xy,yz,zx$ labels the corresponding quadrupolar components. The first two equations enforce the two constraints in Eq.~\eqref{eq:constraints}. The other equations have the form of eigenvalue equations
\begin{equation}
\begin{aligned}
\hat{M}_1(\vec{k})
  \begin{pmatrix}
  S_{T,\vec{k}}/\beta_1\\
  S_{O,\vec{k}}/\beta_2
  \end{pmatrix}
  &=2\lambda_{\vec{k}}^S
  \begin{pmatrix}
  S_{T,\vec{k}}/\beta_1\\
  S_{O,\vec{k}}/\beta_2
  \end{pmatrix},\\
\hat{M}_2(\vec{k})
  \begin{pmatrix}
  Q_{T,\vec{k}}/\beta_1\\
  Q_{O,\vec{k}}/\beta_2
  \end{pmatrix}
  &=2\lambda_{\vec{k}}^Q
  \begin{pmatrix}
  Q_{T,\vec{k}}/\beta_1\\
  Q_{O,\vec{k}}/\beta_2
  \end{pmatrix},  
  \label{eq:eigenvalues}
\end{aligned}
\end{equation}
with matrices $\hat{M}_1$ and $\hat{M}_2$ defined as follows:
\begin{equation}
\begin{aligned}
  \hat{M}_1(\vec{k})=
  &\begin{pmatrix}
  \beta_1^2(J_{2T}-\frac{K_{2T}}{2})g_{\vec{k}} & \beta_1\beta_2(J_{1}-\frac{K_{1}}{2})f_{\vec{k}}\\
  \beta_1\beta_2(J_{1}-\frac{K_{1}}{2})f_{-\vec{k}} & \beta_2^2(J_{2O}-\frac{K_{2O}}{2})g_{\vec{k}}
  \end{pmatrix},\\
  \hat{M}_2(\vec{k})=
  &\begin{pmatrix}
  \beta_3^2\frac{K_{2T}}{2}g_{\vec{k}} & \beta_3\beta_4\frac{K_{1}}{2}f_{\vec{k}}\\
  \beta_3\beta_4\frac{K_{1}}{2}f_{-\vec{k}} & \beta_4^2\frac{K_{2O}}{2}g_{\vec{k}}
  \end{pmatrix},
\end{aligned}
\end{equation}

where
\begin{equation}
\begin{aligned}
    f_{\vec{k}}&=e^{-i(\frac{1}{2}k_x+\frac{\sqrt{3}}{2}k_y)}+e^{ik_x}+e^{-i(\frac{1}{2}k_x-\frac{\sqrt{3}}{2}k_y)}\\
    g_{\vec{k}}&=\cos(\frac{3}{2}k_x-\frac{\sqrt{3}}{2}k_y)
    +\cos\sqrt{3}k_y+\cos(-\frac{3}{2}k_x-\frac{\sqrt{3}}{2}k_y).\nonumber\\
\end{aligned}
\end{equation}
In terms of the eigenvalues in Eq.~\eqref{eq:eigenvalues}, the classical energy becomes
\begin{equation}
    \epsilon(\vec{k},\vec{k}')=\lambda^S_{\vec{k}}\eta\frac{N}{2}(\frac{1}{\beta_1^2}+\frac{1}{\beta_2^2})+\lambda^Q_{\vec{k}'}(\frac{4}{3}-\eta)\frac{N}{2}(\frac{1}{\beta_3^2}+\frac{1}{\beta_4^2})+\epsilon_0,
\end{equation}
where $\epsilon_0$ is a constant that only depends on the coupling constants.

The classical energy $E_\text{class} = \min\limits_{k,k'}(\epsilon)$ is thus determined by minimizing the eigenvalues $\lambda^S_{\vec{k}}$, $\lambda^Q_{\vec{k}'}$ with respect to the ordering wavevectors $\vec{k}$ and $\vec{k}'$ that parametrize the dipolar and quadrupolar spiral order, respectively.
The next step is finding these wavevectors and the constants $\beta_i$ that enter the constraints \eqref{eq:constraints}. 
Without loss of generality, we can set $\beta_1=\beta_3=1$, then $\beta_2$ and $\beta_4$ satisfy the relations $\beta_2=|\psi_{S1}|/|\psi_{S2}|$ and $\beta_4=|\psi_{Q1}|/|\psi_{Q2}|$, expressed in terms of the eigenvectors  $(\psi_{S1},\psi_{S2})$, $(\psi_{Q1},\psi_{Q2})$ of Eq.~(\ref{eq:eigenvalues}).
After considering these two constrains, we finally obtain
\begin{equation}
    \vec{k}=(0,0),~\vec{k}'=(0,0),~\beta_2=0.93,~\beta_4=1.39,
\end{equation}
signalling an intra-unit cell order.
Moreover, we find $\eta=1$, which corresponds to a pure magnetic (dipolar) state. The two components of the eigenvector $(\psi_{S1},\psi_{S2})$ have opposite sign on sites T and O, respectively, which means that the ground state has N\'eel order with the angle $\alpha=180^\circ$ between the two spins. 

We note that while the Luttinger-Tisza method does not allow  to tackle the single-ion anisotropy explicitly, the effect of crystal fields in Eq.~\eqref{eq:anisotropy} with a large positive $\gamma_T \gg J_1 \gg |\gamma_O|$, is only to keep the N\'eel staggered moment  in  the $xy$ plane. It is only once the effect of DM interactions is considered (Section~\ref{sec:MFT}.B) that a noncollinear order with the angle $\alpha \neq 180^\circ$  between the T and O sites develops, as observed experimentally in Ni$_2$Mo$_3$O$_8$.

\section{Crystal field analysis} \label{app:cef}

As discussed in by~\citet{nafexp}, the crystal field plays an important role in Ni$_2$Mo$_3$O$_8$. The previous work used a simple point charge model to determine the crystal field splitting. This model included only the coordinating oxygen ions around each Ni$^{2+}$ ion and used the nominal charge for the neighbour ligands (i.e. -2$|e|$ for O$^{2-}$). The work also showed that without spin-orbit coupling (SOC), the two lowest lying crystal field levels are an orbital singlet $^3\!A$ ground state and an orbital doublet $^3\!E$ excited state. Repeating the calculation, we found that the $^3\!E$ level for the octahedral site is at around 330~meV and does not affect the ground state. The \textit{tetrahedral} site, however, has its $^3\!E$ level at a much lower energy, around 48~meV, which is of the same order of magnitude as the SOC ($\lambda\approx 40$~meV). Thus one should expect the SOC to mix these two orbital levels leading to a large splitting of the orbital singlet (but spin-triplet) $^3\!A$, which is indeed what~\citet{nafexp} found, with a splitting of $\approx23$~meV between the $\Gamma_1$ spin-singlet ground state and $\Gamma_3$ spin-doublet excited state.

The small splitting of the $^{3\!}E$ level also implies that there should be crystal field excitations above this 23~meV level but below 100~meV visible in the neutron spectra (the full calculation implies excitations around 80~meV). However, recent extensive inelastic neutron scattering experiments~\cite{dai_pvt_comm} showed no evidence of this.

Furthermore, the spin-singlet $\Gamma_1$ ground state implies an effective planar single-ion anisotropy with it being highly favourable energetically for the spins to lie in the $ab$ plane. The experimentally determined magnetic structures, however, suggests that the spins on one site is canted by a relatively large angle away from the $ab$ plane. In the case of model 1 (2), this is the octahedral (tetrahedral) site at an angle of $\theta_O-\frac{\pi}{2}$=55$^{\circ}$ ($\theta_T-\frac{\pi}{2}$=34$^{\circ}$). Note that the polar angles $\theta_T$ and $\theta_O$ in Table~\ref{table1} are relative to the $c$-axis.



These experimental findings suggests that the point charge model of~\citet{nafexp} needs some adjustments. In particular, we believe that (1) the splitting between the $^{3\!}E$ and $^{3\!}A$ orbital levels on the tetrahedral site should be much larger, and that (2) the ground state on one of the sites should be the doublet $\Gamma_3$ or a quasi-triplet, rather than the spin-singlet $\Gamma_1$. We can modify the point charge model to satisfy condition 1 by increasing the effective magnitude of the point charges (which increases the magnitude of the crystal field parameters and thus increases the splitting). Condition 2 can be satisfied by including the effects of the Mo$^{4+}$ and Ni$^{2+}$ ions in addition to the O$^{2-}$ in the model, and then either increasing the relative magnitude of the effective charge of the Mo$^{4+}$ ions or decreasing that of the Ni$^{2+}$ ions (even to making it negative) with respects to that of the O$^{2-}$ ions. 

We opted to do both, and posit a point charge model with effective charges which are approximately twice the nominal charges: an effective charge of $-4|e|$ on the oxygen ligands, $+9|e|$ on the molybdenum ligands, and $+1|e|$ on the nickel ligands. The model includes ligands up to $3.5\,\text{\AA}$ away from the magnetic nickel ions, which covers to the nearest molybdenum ligands for each site. This model yields the crystal field parameters shown in Eq.~\eqref{eq:CFT} in the main text. 

This model yields the $^{3\!}A$--$^{3\!}E$ splitting of 95~meV  on the tetrahedral sites and much larger, 950~meV on octahedral sites. The tetrahedral sites still have a $\Gamma_1$ spin-singlet ground state, with the $\Gamma_3$ excited state at $\approx24$~meV and  a further excitation at $\approx125$~meV which may be visible in inelastic neutron scattering data. This structure of the $\Gamma_1-\Gamma_3$ splitting is captured by the relatively large positive value $\gamma_T\approx 30$~meV in the effective spin anisotropy model Eq.~\eqref{eq:anisotropy}.
The octahedral site, on the other hand, has a $\Gamma_3$ spin-doublet ground state with a very low-lying $\Gamma_1$ excited state at $\approx0.5$~meV. In the effective spin anisotropy model Eq.~\eqref{eq:anisotropy} this is reflected in the very small (negative) value of $\gamma_O \approx -0.5$~meV.


Physically, larger magnitudes of the effective charges imply that Ni$_2$Mo$_3$O$_8$ has strong covalent bonds or large charge transfer energies. The larger relative effective charge on the Mo$^{4+}$ ions compared to that on the O$^{2-}$ perhaps reflects the larger extent of the $4d$ orbitals which thus effectively reduces the distance between the magnetic nickel ions and the molybdenum ligand, whilst the smaller relative effective charge on neighbouring Ni$^{2+}$ ions reflects a more itinerant character of the nickel conduction electrons.

Finally, the large difference in the $\Gamma_1$-$\Gamma_3$ splittings for the different sites (octahedral and tetrahedral) in both the original~\cite{nafexp} and our point charge models means that the magnetic excitation spectrum comprises separate bands for the different sites: a low energy magnon-like set of excitations from the octahedral sites, and a higher energy exciton-like set of excitations from the tetrahedral sites. This is indeed seen in the computed magnetic excitation spectra  in Fig.~\ref{fig:spectra}.

\section{Effective single-ion spin anisotropy Hamiltonian} \label{app:cef2}

Due to the three-fold rotation symmetry $C_3$ in the $P6_3mc$ space group, the crystal field Hamiltonian is given by Eq.~\eqref{eq:CFE} in the main text:
\begin{equation}
    \mathcal{H}_{cf}= L_{20}\theta_2\hat{T}_{20} + L_{40}\theta_4\hat{T}_{40} + L_{43}\theta_4\hat{T}_{43}.
\end{equation}
Here $L_{lm}$ are the crystal field parameters whose values in Eq.~\eqref{eq:CFT} were derived from the point charge model,
the coefficients $\theta_l$ are the Stevens factors from \emph{McPhase}~\cite{mcphase} calculation
\begin{equation}
    \theta_2=0.0190,~\theta_4=0.0063.
\end{equation}
and $\hat{T}_{lm}$ are tensorial Stevens--Wybourne operators:
\begin{equation}
\begin{aligned}
\hat{T}_{20}&=\frac{1}{2}(3\hat{L}_z^2-X),\\
\hat{T}_{40}&=\frac{1}{8}(35\hat{L}_z^4-(30X-25)\hat{L}_z^2+3X^2-6X),\\
\hat{T}_{43}&=\frac{\sqrt{35}}{8}((\hat{L}_+^3+\hat{L}_-^3)\hat{L}_z+\hat{L}_z(\hat{L}_+^3+\hat{L}_-^3)),
\end{aligned}
\end{equation}
where $\hat{L}_+$, $\hat{L}_-$ are the ladder operators of the orbital angular momentum and $\hat{L}_z$ is its $z$-component. Here $X=l(l+1)$, which in the present case of Ni$^{2+}$ ($3d^8$) ion with $l=3$ gives $X=12$.\\

Substituting one set of crystal field parameters and the matrix form of operators $\hat{L}_+$, $\hat{L}_-$ and $\hat{L}_z$, the crystal field Hamiltonian becomes a $7\times7$ matrix, and we denote its eigenstates and corresponding eigenvalues  $|n\rangle$ and $E_n$, respectively, with the ground state labeled by $n=0$. Because the value of the coupling constant $\lambda\approx-40$~meV (see Ref.~\onlinecite{soc}) is much smaller than the difference of CFE eigenvalues, 
the spin-orbit coupling $V=\lambda\vec{S}\cdot\vec{L}$ can be treated as a perturbation. The first order of the perturbation is  proportional to $\langle0|\vec{S}\cdot\vec{L}|0\rangle$, which vanishes identically. The correction to the energy in the second order perturbation theory is of the form
\begin{equation}
    E_0^{2nd}=\sum_{m>0}\frac{|\langle0|V|m\rangle|^2}{E_0-E_m}=\lambda^2\sum_{ij}\Lambda_{ij}S^iS^j,
\end{equation}
where
\begin{equation}
  \Lambda_{ij}=\sum_{m>0}\frac{\langle0|L_i|m\rangle\langle m|L_j|0\rangle}{E_0-E_m} 
\end{equation}
is the single-ion anisotropy parameter. Finally we obtain the effective single-ion spin anisotropy Hamiltonian in Eq.~\eqref{eq:anisotropy} of the main text:
\begin{equation}
\label{eq:anisotropy_app}
    \mathcal{H}_\text{A}=\sum_{T}\gamma_{T}(S_i^z)^2+\sum_{O}\gamma_{O}(S_i^z)^2
\end{equation}
with numerical values of the coefficients $\gamma_{T}=30.41$ meV, $\gamma_{O}=-0.53$ meV.

\section{LSWT calculation}\label{app:Spinwave}

The full Hamiltonian of the effective Heisenberg model for the N\'eel state consists of the exchange interactions in Eq.~\eqref{eq:effneelmodel}, with the addition of the DM interaction and the single-ion anisotropy:
\begin{equation}\label{eq:Hneel-full}
\begin{aligned}
    \mathcal{H}_{\text{N\'eel}}^{\text{e}}&=\sum_{\langle ij\rangle}J_{1}^{\text{e}}\vec{S}_i\cdot\vec{S}_j
    +\sum_{\langle ij\rangle}\vec{D}_{ij}\cdot(\vec{S}_i\times\vec{S}_j)\\
    &+\sum_{\langle\langle ij\rangle\rangle,T}J_{2T}^{\text{e}}\vec{S}_i\cdot\vec{S}_j
    +\sum_{\langle\langle ij\rangle\rangle,O}J_{2O}^{\text{e}}\vec{S}_i\cdot\vec{S}_j
    \\
&+\sum_{i,T}\gamma_T(S_i^z)^2+\sum_{i,O}\gamma_O(S_i^z)^2.
\end{aligned}
\end{equation}
The effective coupling constants $J_{1}^{\text{e}}$, $J_{2T}^{\text{e}}$ and  $J_{2O}^{\text{e}}$ are quoted in Eq.~\eqref{eq:effneelparameter} in the main text.
%
After the Holstein-Primakoff transformation in Eq.~(\ref{eq:HPtrans}), the Hamiltonian becomes 
\begin{equation}
    \mathcal{H}_{\text{N\'eel}}^{\text{e}}=\mathcal{E}_{\text{N\'eel}}+\mathcal{E}_{C}+\sum_{\vec{k}}\psi_{\vec{k}}^{\dagger}H(\vec{k})\psi_{\vec{k}},
\end{equation}
where $\mathcal{E}_{\text{N\'eel}}$ is the MF energy expression, $\mathcal{E}_C$ is a constant term originating from the commutation relation when we construct the bosonic Nambu representation. Here the composite vector $\psi_{\vec{k}}=[a({\vec{k}}),b({\vec{k}}),{a^{\dagger}}({-\vec{k}}),{b^{\dagger}}({-\vec{k}})]^{\intercal}$ consists of the bosonic operators $a$ on T site and operators $b$ on O sites. The matrix $H(\vec{k})$ is
\begin{equation}\label{eq:Hmatrix-Neel}
\begin{aligned}
H(\vec{k})&=
\begin{pmatrix}
f_T(\vec{k}) & 0 & \frac{\gamma_T}{2} & g(\vec{k})\\
0 & f_O(\vec{k}) & g(-\vec{k}) & \frac{\gamma_O}{2}\\
\frac{\gamma_T}{2} & g(-\vec{k})^* & f_T(\vec{k}) & 0\\
g(\vec{k})^* & \frac{\gamma_O}{2} & 0 & f_O(-\vec{k})\\
\end{pmatrix},\\
\end{aligned}
\end{equation}
where
\begin{equation}
\label{eq:HMneel}
\begin{aligned}
    f_T(\vec{k})&=\frac{3}{2}J_1^e-3J_{2T}^e+\frac{1}{2}\gamma_T+J_{2T}^e\Big(\cos(\frac{3}{2}k_x-\frac{\sqrt{3}}{2}k_y\\
    &+\cos\sqrt{3}k_y+\cos(-\frac{3}{2}k_x-\frac{\sqrt{3}}{2}k_y)\Big),\\
     f_O(\vec{k})&=\frac{3}{2}J_1^e-3J_{2O}^e+\frac{1}{2}\gamma_O+J_{2O}^e\Big(\cos(\frac{3}{2}k_x-\frac{\sqrt{3}}{2}k_y)\\
    &+\cos\sqrt{3}k_y+\cos(-\frac{3}{2}k_x-\frac{\sqrt{3}}{2}k_y)\Big),\\
    g(\vec{k})&=-\frac{1}{2}J_1^e(e^{-i(\frac{1}{2}k_x+\frac{\sqrt{3}}{2}k_y)}+e^{ik_x}+e^{-i(\frac{1}{2}k_x-\frac{\sqrt{3}}{2}k_y)})\\
    &-iD\Big(e^{-i(\frac{1}{2}k_x+\frac{\sqrt{3}}{2}k_y)}\sin(\phi+\frac{2\pi}{3})+e^{ik_x}\sin\phi\\
    &+e^{-i(\frac{1}{2}k_x-\frac{\sqrt{3}}{2}k_y)}\sin(\phi+\frac{4\pi}{3})\Big).
\end{aligned}    
\end{equation}
Above, $\phi$ is the asimuthal angle of the spin direction in $xy$ plane. 
The constant  $\mathcal{E}_C$ in Eq.~\eqref{eq:Hmatrix-Neel} is given by
\begin{equation}
    \mathcal{E}_C=-\frac{3\sqrt{3}}{4(2\pi)^2}\int_{BZ}d^2k\Big(f_T(\vec{k})+f_O(\vec{k})\Big).
\end{equation}
After the Bogoliubov transformation, the LSWT energy of N\'eel state becomes
\begin{equation}
    \mathcal{E}_{\text{N\'eel}}^{LSWT}=\mathcal{E}_{\text{N\'eel}}+\mathcal{E}_{C}+\frac{3\sqrt{3}}{4(2\pi)^2}\int_{1BZ}\Big(E_a(\vec{k})+E_b(\vec{k})\Big),
\end{equation}
where $E_a(\vk)$ and $E_b(\vk)$ are the two positive eigenvalues of the matrix Eq.~\eqref{eq:Hmatrix-Neel}, corresponding physically to the two bands in the magnetic spectrum.
Having fixed the exchange couplings to their 
\textit{ab initio} values in Eq.~(\ref{eq:DFT}) and the single-ion anisotropy parameters as quoted below Eq.~\eqref{eq:anisotropy}, our calculations show that the N\'eel state has the minimum energy for the asimuthal angle $\phi=30^{\circ}$.\\ 

The full Hamiltonian of the effective Heisenberg model for the NCAF state is given by the spin bilinears in Eqs.~\eqref{eq:effncafmodel}-\eqref{eq:effncafparameter} in the main text, with the addition of the DM interaction and single-ion spin anisotropy terms, analogous to the Eq.~\eqref{eq:Hneel-full} above.

After the Holstein-Primakoff transformation in Eq.~(\ref{eq:HPtrans}), the Hamiltonian becomes
\begin{equation}\label{HSW}
    \mathcal{H}=\mathcal{E}_{\text{NCAF}}+\mathcal{E}_{C}+H_\text{lin}[b^{\dagger},\,b]+\psi_{\vec{k}}^{\dagger}H(\vec{k})\psi_{\vec{k}},
\end{equation}
where $\mathcal{E}_{\text{NCAF}}$ is the mean field energy and $H_\text{lin}[b^{\dagger},b]$ collects the terms linear in the boson creation and annihilation operators (these terms are ignored in what follows as they do not conserve the magnon number). 

The composite vector $\psi_{\vec{k}}$ of creation-annihilation operators is given by
\begin{equation}\label{eq:vector}
\psi=[a({\vec{k}}),{a^{\dagger}}({-\vec{k}}),b({\vec{k}}),{b^{\dagger}}({-\vec{k}}),c({\vec{k}}),{c^{\dagger}}({-\vec{k}}),d({\vec{k}}),{d^{\dagger}}({-\vec{k}})]^{\intercal},
\end{equation}
where $a$ and $c$ are annihilation operators at T and -T sites, whereas $b$ and $d$ annihilate bosons on the O and -O sites (see Fig.~\ref{fig:approxH}b for the notation of the sites). The matrix $H(\vec{k})$ in Eq.~\eqref{HSW} is given by
\begin{equation}\label{eq:Hmatrix-NCAF}
H(\vec{k},\vec{\theta})=\frac{1}{2}
\begin{pmatrix}
H_{11}(\vec{k},\vec{\theta}) & H_{12}(\vec{k},\vec{\theta})\\
H_{12}^{\dagger}(\vec{k},\vec{\theta}) & H_{11}^{\intercal}(-\vec{k},\vec{\theta})\\
\end{pmatrix}
\end{equation}
with the entries
\begin{equation}
\begin{aligned}
H_{11}(\vec{k})&=
\begin{pmatrix}
f_a(\vec{k},\vec{\theta}) & 0 &f_{ab}(\vec{k},\vec{\theta}) & f_{ad}(\vec{k},\vec{\theta})\\
0 & f_c(\vec{k},\vec{\theta}) &f_{cb}(\vec{k},\vec{\theta}) & f_{cd}(\vec{k},\vec{\theta})\\
f_{ab}^*(\vec{k},\vec{\theta}) & f_{cb}^*(\vec{k},\vec{\theta}) & f_b(\vec{k},\vec{\theta}) & 0\\
f_{ad}^*(\vec{k},\vec{\theta}) & f_{cd}^*(\vec{k},\vec{\theta}) & 0 & f_d(\vec{k},\vec{\theta})\\
\end{pmatrix}\\
H_{12}(\vec{k})&=
\begin{pmatrix}
g_a(\vec{k},\vec{\theta}) & g_{ac}(\vec{k},\vec{\theta}) & g_{ab}(\vec{k},\vec{\theta}) & g_{ad}(\vec{k},\vec{\theta})\\
g_{ac}(-\vec{k},\vec{\theta}) & g_c(\vec{k},\vec{\theta}) & g_{cb}(\vec{k},\vec{\theta}) & g_{cd}(\vec{k},\vec{\theta})\\
g_{ab}(-\vec{k},\vec{\theta}) & g_{cb}(-\vec{k},\vec{\theta}) & g_b(-\vec{k},\vec{\theta}) & g_{bd}(\vec{k},\vec{\theta})\\
g_{ad}(-\vec{k},\vec{\theta}) & g_{cd}(-\vec{k},\vec{\theta}) & g_{bd}(-\vec{k},\vec{\theta}) & g_d(\vec{k},\vec{\theta})\\
\end{pmatrix}\nonumber\\
\end{aligned}
\end{equation}
where $\vec{\theta}=(\theta_O,\theta_T)$, and the matrix elements are given by a lengthy set of expressions shown here for completeness:
\beq
\begin{aligned}
f_a(\vec{k},\theta_T,\theta_O)&=-2J_{1+}\cos(\theta_T+\theta_O)+J_{1-}\cos(\theta_T+\theta_O)\\
&-2J_{TF}+4J_{TA}\\
&+2J_{TF}\cos{\sqrt{3}k_y}+2D\sin(\theta_T+\theta_O)\\
&+\gamma_T(\sin^2\theta_T-2\cos^2\theta_T),\\
f_b(\vec{k},\theta_T,\theta_O)&=-2J_{1+}\cos(\theta_T+\theta_O)+J_{1-}\cos(\theta_T+\theta_O)\\
&-2J_{OF}+4J_{OA}\\
&+2J_{OF}\cos{\sqrt{3}k_y}+2D\sin(\theta_T+\theta_O)\\
&+\gamma_O(\sin^2\theta_T-2\cos^2\theta_T),\\
f_c(\vec{k},\theta_T,\theta_O)&=f_a(\vec{k},\theta_T,\theta_O),\\
f_d(\vec{k},\theta_T,\theta_O)&=f_b(\vec{k},\theta_T,\theta_O),
\end{aligned}
\eeq
\begin{equation}\label{eq:Hmatrixele1}
\begin{aligned}
f_{ab}(\vec{k},\theta_T,\theta_O)&=\Big(\frac{J_{1+}(\vec{\theta})}{2}(\cos(\theta_T+\theta_O)+1)\\
&-\frac{D}{4}\sin(\theta_T+\theta_O)\\
&+\frac{\sqrt{3}}{4i}D(\sin\theta_O-\sin\theta_T)\Big)e^{-i(\frac{1}{2}k_x+\frac{\sqrt{3}}{2}k_y)}\\
&+\Big(\frac{J_{1+}(\vec{\theta})}{2}(\cos(\theta_T+\theta_O)+1)\\
&-\frac{D}{4}\sin(\theta_T+\theta_O)\\
&-\frac{\sqrt{3}}{4i}D(\sin\theta_O-\sin\theta_T)\Big)e^{-i(\frac{1}{2}k_x-\frac{\sqrt{3}}{2}k_y)},\\
f_{ad}(\vec{k},\theta_T,\theta_O)&=\Big(\frac{J_{1-}(\vec{\theta})}{2}(1-\cos(\theta_T+\theta_O))\\
&-\frac{D}{2}\sin(\theta_T+\theta_O)\Big)e^{ik_x},\\
f_{cb}(\vec{k},\theta_T,\theta_O)&=f_{ad}(\vec{k},\theta_T+\pi,\theta_O+\pi),\\
f_{cd}(\vec{k},\theta_T,\theta_O)&=f_{ab}(\vec{k},\theta_T+\pi,\theta_O+\pi),\\
\end{aligned}
\end{equation}

\begin{equation}\label{eq:Hmatrixele2}
\begin{aligned}
g_a(\vec{k},\theta_T,\theta_O)&=g_c(\vec{k},\theta_T,\theta_O)=\gamma_T\sin^2\theta_T,\\
g_b(\vec{k},\theta_T,\theta_O)&=g_d(\vec{k},\theta_T,\theta_O)=\gamma_O\sin^2\theta_O,\\
g_{ac}(\vec{k},\theta_T,\theta_O)&=-4J_{TA}\cos{\frac{3k_x}{2}}\cos{\frac{\sqrt{3}k_y}{2}}\\
g_{bd}(\vec{k},\theta_T,\theta_O)&=-4J_{OA}\cos{\frac{3k_x}{2}}\cos{\frac{\sqrt{3}k_y}{2}}\\
g_{ab}(\vec{k},\theta_T,\theta_O)&=\Big(\frac{J_{1+}(\vec{\theta})}{2}(\cos(\theta_T+\theta_O)-1)\\
&-\frac{D}{4}\sin(\theta_T+\theta_O)\\
&+\frac{\sqrt{3}}{4i}D(\sin\theta_O+\sin\theta_T)\Big)e^{-i(\frac{1}{2}k_x+\frac{\sqrt{3}}{2}k_y)}\\
&+\Big(\frac{J_{1+}(\vec{\theta})}{2}(\cos(\theta_T+\theta_O)-1)\\
&-\frac{D}{4}\sin(\theta_T+\theta_O)\\
&-\frac{\sqrt{3}}{4i}D(\sin\theta_O+\sin\theta_T)\Big)e^{-i(\frac{1}{2}k_x-\frac{\sqrt{3}}{2}k_y)},\\
g_{ad}(\vec{k},\theta_T,\theta_O)&=\Big(-\frac{J_{1-}(\vec{\theta})}{2}(1+\cos(\theta_T+\theta_O))\\
&-\frac{D}{2}\sin(\theta_T+\theta_O)\Big)e^{ik_x},\\
g_{cb}(\vec{k},\theta_T,\theta_O)&=g_{ad}(\vec{k},\theta_T+\pi,\theta_O+\pi),\\
g_{cd}(\vec{k},\theta_T,\theta_O)&=g_{ab}(\vec{k},\theta_T+\pi,\theta_O+\pi).\\
\end{aligned}
\end{equation}

This results in the constant contribution to the energy $\mathcal{E}_C$:
\begin{equation}
    \mathcal{E}_C=-\frac{3\sqrt{3}}{4(2\pi)^2}\int_{BZ}d^2k\Big(f_a(\vec{k})+f_b(\vec{k})+f_c(\vec{k})+f_d(\vec{k})\Big).
\end{equation}
After the Bogoliubov transformation, the energy of the NCAF state within the linear spin-wave theory is finally given by 
\begin{eqnarray}
    \mathcal{E}_{\text{NCAF}}^{LSWT}&=&\mathcal{E}_{\text{NCAF}}+\mathcal{E}_{C}\\
    &+&\frac{3\sqrt{3}}{4(2\pi)^2}\int_{1BZ}(E_a(\vec{k})+E_b(\vec{k})+E_c(\vec{k})+E_d(\vec{k})).\nonumber
\end{eqnarray}
The resulting energy of the NCAF state is computed by summing over the real, positive eigenvalues of the matrix $H(\vk,\theta_T,\alpha)$ in Eq.~\eqref{eq:Hmatrix-NCAF} and optimizing the angles $\alpha$ and $\theta_{T}$ such as to minimize this energy. The resulting optimal values of $\alpha$ are shown in Fig.~\ref{phasediagram5} in the main text as a function of the varying DM interaction strength.

\section{Magnetic excitation spectrum} \label{app:Spinwave_spectrum}

In linear spin wave theory (LSWT) the single-ion anisotropy (SIA) is a constant term added to the diagonal of the Hamiltonian matrix. The resulting magnetic spectrum is shown in Fig.~\ref{fig:spectra}(a) in the main text.
On the other hand, in the RPA calculations the full single-ion Hamiltonian matrix is calculated first and diagonalised, then an RPA coupling is made for each dipolar transition between the single-ion states. Given the large difference in the SIA between the tetrahedral (large gap) and octahedral (small gap) sites, the result is that in the RPA calculation, despite the large nearest neighbour $J_1$ and $K_1$ interactions there is little coupling between the modes associated with the tetrahedral and octahedral sites. In contrast, in LSWT the additional diagonal constants in the Hamiltonian serve to separate out the tetrahedral and octahedral modes in energy but the off-diagonal terms in the Hamiltonian still results in significant coupling between the sites which thus modifies the dispersion. One can see the resulting differences by comparing the magentic spectra in the two panels in Fig.~\ref{fig:spectra}.
Given the relatively large SIA, we believe that its treatment in LSWT is less accurate than with the RPA but this should be confirmed by experimental measurements of the magnetic excitation spectrum.


\bibliography{cite}

\end{document}